$Pb_{0.95}La_{0.05}Zr_{0.54}Ti_{0.46}O_3$ THIN FILMS FOR PHOTOVOLTAIC APPLICATIONS

by

HARSHAN VASUDEVAN NAMPOORI

SUSHMA KOTRU, COMMITTEE CHAIR
DAWEN LI
PATRICK KUNG
JABER ABU QAHOUQ
SHANLIN PAN

A DISSERTATION

Submitted in partial fulfillment of the requirements
for the degree of Doctor of Philosophy in the
Department of Electrical and Computer Engineering
in the Graduate School of
The University of Alabama

TUSCALOOSA, ALABAMA

2012

ABSTRACT

Ferroelectrics have shown potential as a promising alternative material for future photovoltaic applications. Observance of high open circuit voltages in ferroelectric thin films, have generated considerable interest in the field of ferroelectric photovoltaic in recent years. The field of ferroelectric photovoltaic is evolving and not yet completely understood compared to the semiconductor based photovoltaic technology. This dissertation investigates photovoltaic characteristics of ferroelectric films in an attempt to further the understanding of ferroelectric photovoltaic.

This dissertation presents photovoltaic properties of ferroelectric $Pb_{0.95}La_{0.05}Zr_{0.54}Ti_{0.46}O_3$ thin films. The films were fabricated by solution based method and spin coating technique. The post annealing process on these films was optimized to achieve the desired ferroelectric and dielectric properties. A measurement setup was established to study the PV characteristic of the devices. Dependence of current-voltage (I-V) behavior of the cells on parameters such as electrical poling, annealing temperature, nature of top electrodes, and intensity of illumination, were investigated. The photovoltaic response was shown to improve by using electrodes with low work functions. An electric circuit model was developed to simulate the behavior of a single ferroelectric photovoltaic cell and the dependence of open circuit voltage ($V_{oc}$) and short circuit current ($I_{sc}$) on light intensity.

# DEDICATION

To my family.

## LIST OF ABBREVIATIONS AND SYMBOLS

| | |
|---|---|
| *A* | Area of the capacitor |
| A* | Richardson constant |
| AFM | Atomic force microscope |
| *C* | Capacitance |
| CSD | Chemical solution deposition |
| *D* | Thickness of film |
| *E* or $E_k$ | Electrical field |
| $e_{31,f}$ | Effective transverse piezoelectric coefficient for thin film |
| $E_c$ | Coercivity |
| $E_{dp}$ | Depolarization field |
| $E_i$ | Interface electric field |
| FeRAM | Ferroelectric random access memory |
| FP | Frenkel poole emission |
| G | Irradiance level |
| *H* | Hole |
| *I* | Electrical current |
| $I_{sc}$ | Short circuit current |
| ITO | Indium tin oxide |
| I-V | Current- voltage |
| *J* | Current density |
| MEMS | Microelectromechanical system |
| MOD | Metal-organic decomposition |
| MPB | Morphotropic phase boundary |
| N | Carrier concentration |
| *n* | Refractive index |
| $N_{dop}$ | Doping density |
| P | Polarization |
| PED | Pulsed electron depostion |
| PLD | Pulsed laser deposition |
| PLZT | La-doped lead zirconate titanate |
| PLZT | $Pb_{0.95}La_{0.05}Zr_{0.46}Ti_{0.54}O_3$ |
| $P_r$ | Remanent polarization |
| $P_s$ | Spontaneous polarization |
| PV | Photo voltaic |
| PZT | Lead zirconate titanate |
| *Q* | Charge |
| R | Coefficient of determination |
| RMS | Root mean square |
| RPM | Round per minute |
| RTA | Rapid thermal annealing |
| SCLC | Space charge-limited current |
| T | Temperature |
| tan$\delta$ | Dissipation factor |
| TCO | Transparent conducting oxide |
| UV | Ultra violet |
| *V* | Voltage |
| VCCS | Voltage controlled current source |
| $V_{oc}$ | Open circuit voltage |
| XPS | X-ray photoelectron spectroscopy |
| XRD | X-ray diffractometer |
| $\mu$ | Mobility |
| $\varepsilon_0$ | Permittivity of vacuum |
| $\rho$ | Resistivity |
| $\varphi_B$ | Work function  barrier height |

# ACKNOWLEDGMENTS

I would like to express my sincere gratitude and appreciation to Dr. Sushma Kotru, my advisor whose wholehearted encouragement and patience helped me to complete this dissertation.

This dissertation would not have been possible without financial support from multiple sources. I would like to acknowledge, Department of Electrical and Computer Engineering, and MINT center, University of Alabama, NSF Grant EECS 943711, and Dr. Subhadra Gupta for support.

I am grateful to members of the committee, Dr. Dawen Li, Dr. Jaber Abu Qahouq, Dr.Patrick Kung, and Dr. Shanlin Pan their constructive suggestions for this dissertation. I would also like to acknowledge Dr. Susan L Burkett and Dr. Arunava Gupta for their input and interests for this work. This research became possible through the use of characterization facilities at MINT Center, Central Analytical Facility (CAF) and Micro fabrication Facility (MFF) at the University of Alabama. I would like to thank staff of MINT, CAF and ECE for their kind co-operation.

It would be impossible for me to forget the love, care and consideration of all my friends who made my stay pleasant in the campus and the life outside. I would also like to thank my family and my wife Sruthi, for standing with me throughout the trials and tribulations of the graduate study.

CONTENTS

LIST OF TABLES

LIST OF FIGURES

# CHAPTER 1

# INTRODUCTION

## 1.1 Overview of the Research Project

The objective of this research is to demonstrate the photovoltaic response of thin films of a ferroelectric material and evaluate their use for future photovoltaic applications such as solar cells and/or UV sensors. The ferroelectric material chosen for this study was Lanthanum doped Lead Zirconate Titanate represented by the chemical formulae $Pb_{0.95}La_{0.05}Zr_{0.54}Ti_{0.46}O_3$ (PLZT). Solution based processing, commonly known as chemical solution deposition, was used to fabricate these films [1-2]. This method utilizes metal organic precursors to make chemical solutions of a desired composition. For this work, the chemical solutions were prepared in-house using the Metal Organic Decomposition (MOD) route. This route has been extensively used to grow high quality ferroelectric thin films of various materials [1-3]. The advantage of this technique is that films can be prepared at room temperature at ambient atmospheric conditions, thereby alleviating the need for high vacuum processes. The method is inexpensive and scalable to larger substrate sizes. To achieve the objective, the following tasks were accomplished:

1. Optimize the processing conditions to achieve high quality films of PLZT.

2. Characterize the grown films for phase and ferroelectric properties.

3. Fabricate test solar cells in capacitor configuration and measure their photovoltaic response.

4. Use innovative methods to improve the photovoltaic characteristics of these test devices.

5. Predict the behavior of the devices using mathematical simulation model.

## 1.2 Photovoltaic Technology

Photovoltaic (PV) materials and devices convert sunlight into electrical energy, and photovoltaic cells are commonly known as solar cells. Photovoltaic technology is the application of photovoltaic cells for generating electricity from light. One of the attractions of PV technology is the ability to generate electricity in a clean, quiet and reliable way. The current market of PV technology is dominated by semiconductor-based solar cells. In a conventional semiconductor-based PV cell, when illuminated, charge carriers (electron–hole pairs) are generated at the depletion region (p-n junction) of the semiconductor. These generated carriers give rise to photo current and photo voltage, which can be harvested in a useful manner. The voltage produced depends on the electronic band gap of the semiconductor. The PV technologies differ on the basis of the type of material used, which determines the current voltage (I-V) characteristics and the efficiency of the photovoltaic device. Figure 1.1 shows development of the PV cell technology over time and the conversion efficiency of these devices [4].

Photovoltaic cells, based on silicon and III-V semiconductor compounds, have high conversion efficiencies. Single junction and multi junction derivatives of these materials have the highest conversion efficiencies and are commonly used in space and terrestrial applications. Organic, inorganic and quantum dot cells are other emerging PV cell technologies; though the efficiencies of these cells are lower in comparison to their semiconductor counterparts, they have found a place in figure 1.1 due to their cost effectiveness.

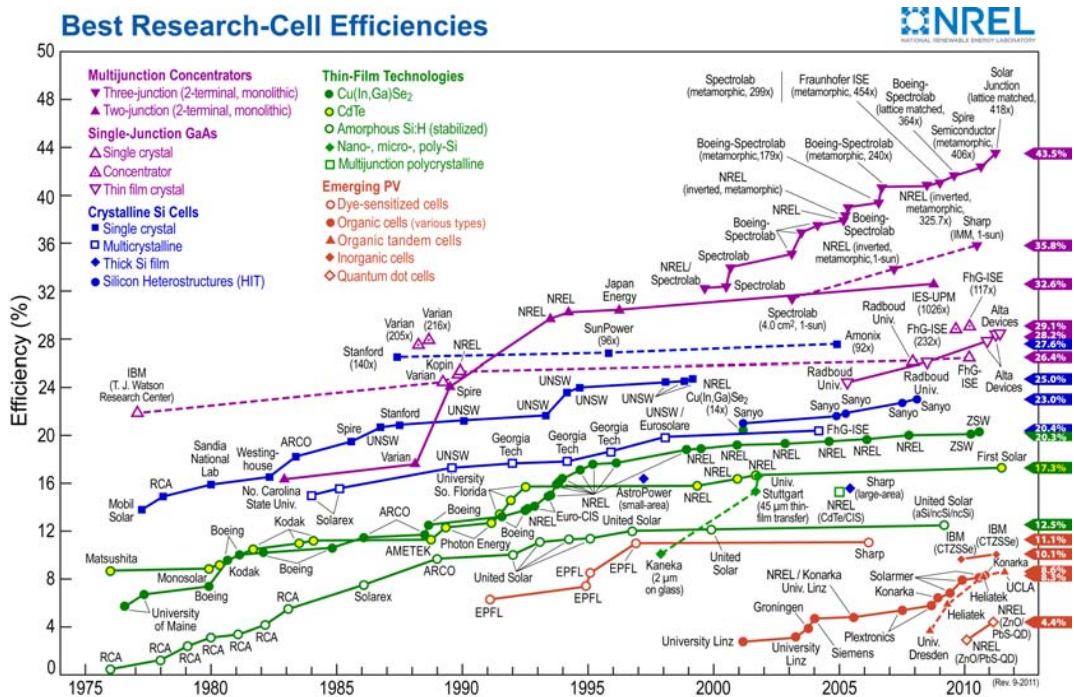

**Figure 1.1** Current status of PV cells and their efficiency [4].

Solar cells have achieved an efficiency of ~ 43.5 % as a result of research and development efforts in the last 30 years [4]. There is an ongoing quest to improve the efficiency of the PV cells further, and to reduce the fabrication cost. Generation of maximum power per dollar is the main factor, which drives the PV research and the industry forward. Exploring the viability of alternate materials for increasing the efficiency of PV cells has gained momentum in the past decade. Ferroelectric materials are one of the latest additions among these, and this research area is generally referred to as ferroelectric photovoltaic.

## 1.3   Ferroelectric Photovoltaic

### 1.3.1   Ferroelectric Materials

In a crystal, the atoms are arranged in a periodic pattern. The smallest unit which defines the periodic pattern in the crystal is called a unit cell. The crystallinity of the solid arises from spatial arrangement of atoms in the unit cells, which are held together by inter-atomic bonds. In some materials, due to the nature of these inter-atomic bonds and the arrangement of atoms, the centers of positive and the negative charge in the unit cell coincide. This leads to the formation of centro-symmetric unit cells. However, there are other types of unit cells, which are not centro-symmetric, in which spontaneous polarization can exist. The spontaneous polarization in the crystal produces an electrical dipole moment which is affected by an external electric field. While this electric field is applied to the non centro-symmetric materials, it can control the electrical dipole moments causing the polarization to increase, decrease or reverse the direction. The property of certain materials to exhibit a spontaneous electric polarization, which can be reversed by application of an external electric field, is called ferroelectricity [5].

Crystal structure can be classified into 32 point groups. Based on their symmetries, 21 of them are non centro-symmetric and 11 are centro-symmetric. The 20 out of these 21 non-centro-symmetric groups are piezoelectric, and 10 of these 20 piezoelectric groups are pyroelectric. Piezoelectricity is the induced electrical charge in the material, in response to an applied mechanical stress. Pyroelectricity is the ability of the materials to generate voltage or current in response to change in temperature. All ferroelectric materials are piezoelectric as well as pyroelectric. Typical examples of ferroelectric materials are Barium Titanate ($BaTiO_3$), Potassium Niobate ($KNbO_3$), Barium Strontium Titanate ($Ba_xSr_{1-x}$)$TiO_3$ (BST) and Lead Zirconate Titanate $Pb(Zr_xTi_{1-x})O_3$ commonly referred as PZT.

In this work, Lanthanum doped PZT was used to explore the photovoltaic response. Lead zirconnate titanate (PZT), a solid solution of $PbZrO_3$ and $PbTiO_3$, is one of the most extensively studied material system at the morphotropic phase boundary (MPB) with composition of Zr/Ti ration 52/48. At the MPB composition, the tetragonal and rhombohedral phases of PZT co-exists, giving rise to high ferroelectric and piezoelectric values. These properties can also be tuned by doping with other elements. Doping with Nb is reported to increase the effective transverse piezoelectric coefficient -$e_{31,f}$ as well as pyroelectric response [6]. Similarly, doping PZT with an appropriate amount of La has shown to enhance the optical properties of the material.

Unit cell structures of PZT ferroelectric material with two polarization states (up and down) are shown in figure 1.2 (a) and (b). A shift in the center ions of $Zr^{4+}$/ $Ti^{4+}$ ions with respect to the oxygen tetrahedron correspondingly leads to an up polarization state and a down polarization state. An electric field, when applied to the up polarization state of the crystal, can shift the $Zr^{4+}$/ $Ti^{4+}$ ions and the oxygen tetrahedron in the opposite direction, causing polarization switching. A few adjacent ferroelectric unit crystals of similar polarization align together, to form a ferroelectric domain.

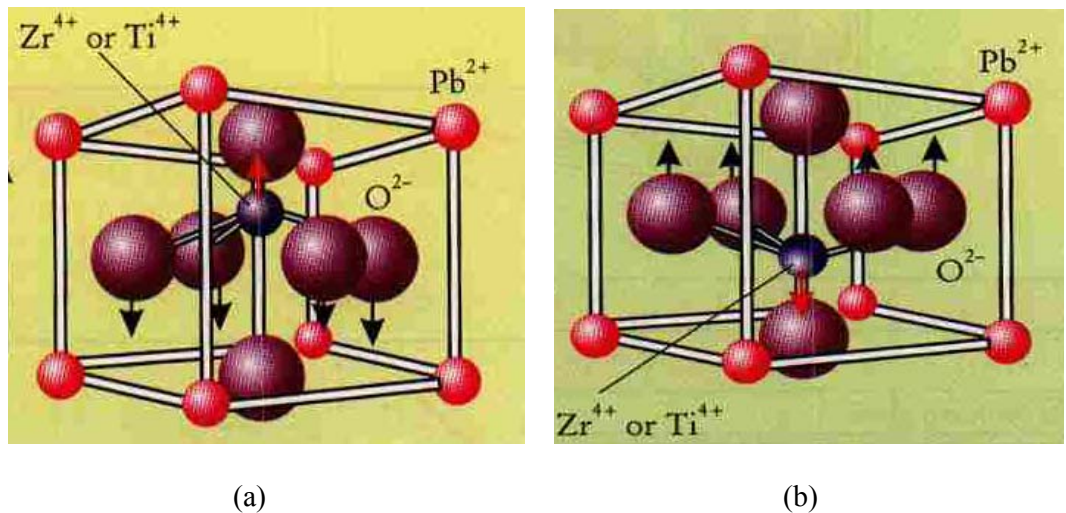

(a)                                              (b)

**Figure 1.2** Unit cell structure of a PZT cubic perovskite with (a) up polarized (b) down polarized crystal structure [7]

Ferroelectric material exhibits a hysteresis behavior with applied electric field as shown in figure 1.3 (a). As shown in the figure, with an increase of the applied electric field, the numbers of domains grow and orient themselves with the applied field, gradually till they attain the saturation polarized state ($P_s$). For each ferroelectric domain, the polarization points towards a particular orientation. With removal of the electric field most of the domains do not return to the random aligned states. Some of these domains retain a stable state of polarization, which is called as the remanent polarization ($P_r$) state of the ferroelectric. Figure 1.3 (b) also shows corresponding ferroelectric domain orientations in the crystal, with the applied electric field at each point of the hysteresis curve. The remanent polarization points to +$P_r$ (at zero electric field) after the crystal is subjected to a positive electric field, or switches to –$P_r$ state with an application of negative field. Thus $P_r$ can be switched with application of electric field. The ability to switch the polarization with an applied electric field makes the ferroelectric materials attractive towards solid state memories (FeRAM) as well as photovoltaic applications.

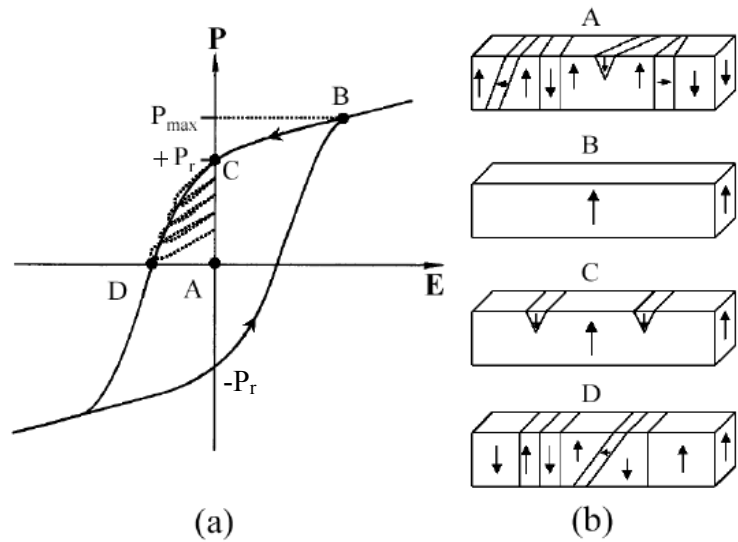

**Figure 1.3** A typical (a) ferroelectric hysteresis loop with (b) corresponding domain alignments [8].

### 1.3.2  Mechanism of Ferroelectric Photovoltaic

When a ferroelectric material is illuminated with light of wavelength corresponding to the energy band gap ($E_g$) of the material, charge carriers (electron-hole pairs) are generated. These photo generated carriers are separated and driven to the electrodes by the polarization induced internal electric field, causing a photovoltaic output. For a junction-based semiconductor photovoltaic device, the electric field, which exists at the depletion layer at the interface (p-n junction), separates the charge carriers. Thus, photovoltaic effect in ferroelectrics is a bulk-based effect, which differs from the junction-based semiconductor photovoltaic effect. Since the internal electric field is not limited to an interfacial region in a ferroelectric, PV responses can be generated without forming complex junction structures. A simplified schematic of this mechanism is shown in figure 1.4 (a and b).

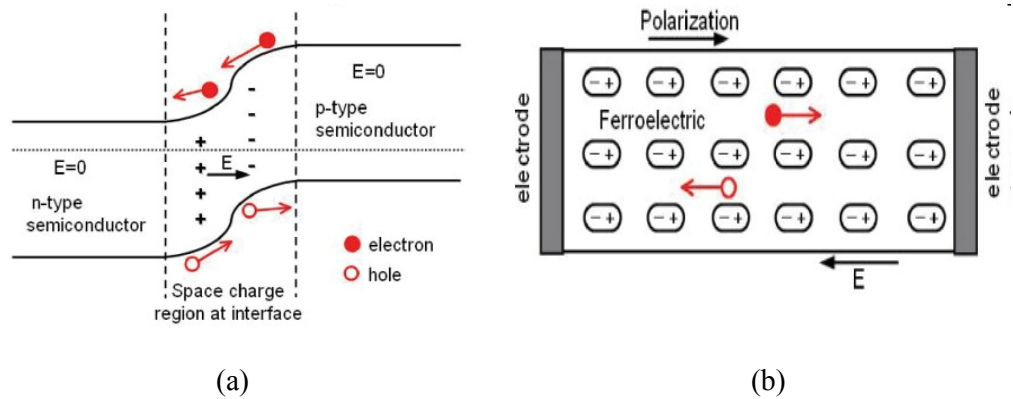

(a) (b)

**Figure 1.4** Simplified schematics of PV mechanism in (a) semiconductor p-n junction and (b) ferroelectric thin film [9].

### 1.3.3 History and Current Status of Research

The photovoltaic effect in ferroelectric single crystals and ceramics (bulk) such as $BaTiO_3$, $Pb(Zr,Ti)O_3$, and $LiNbO_3$ was observed earlier [10-12]. The origin of this effect was attributed to the non-centrosymmetric nature of the unit cell [13,14]. The photovoltaic efficiency in these crystal/bulk materials was, however, limited due to small current densities of the order of nano-amperes/cm$^2$ owing to their large band gaps.

The renaissance of research in photovoltaic ferroelectric materials (FE-PV) was a consequence of the observation of photo voltage, larger than the band gap of the material, in thin films of Bismuth ferrite ($BiFeO_3$). An open circuit voltage ($V_{oc}$) of ~20V was obtained from $BiFeO_3$ thin films, which have a band gap ($E_g$) of 2.67 eV [15,16]. It was also shown that the photo voltage could be reversed or enhanced, by controlling the ferroelectric polarization which was in turn controlled by electrical poling on lead based thin films [19]. Poling is the process of applying an electric voltage higher than the coercive field, to a ferroelectric material, while cooling it from transition temperature to room temperature. Poling helps to orient ferroelectric domains in one particular direction, resulting in maximum polarization.

PLZT is another class of ferroelectric material which is being studied for photovoltaic properties. One of the early works investigating the photovoltaic responses was based on ceramic of $Pb_{0.97}La_{0.03}Zr_{0.52}Ti_{0.48}O_3$ (3/52/48) composition [17]. The photo response of various other compositions of PLZT ceramics was studied extensively by Poosanaas et al [18]. The maximum photocurrent was obtained from $Pb_{0.97}La_{0.03}Zr_{0.52}Ti_{0.48}O_3$ (3/52/48) and the maximum photo voltage was obtained from $Pb_{0.95}La_{0.05}Zr_{0.54}Ti_{0.46}O_3$ (5/54/46), as depicted in figure 1.4 [18].

It was Qin et al., who investigated the photovoltaic properties of PLZT thin films [19-22]. Their work was based on $Pb_{0.97}La_{0.03}Zr_{0.52}Ti_{0.48}O_3$ (3/52/48). They reported that power conversion efficiency (~0.01%) on epitaxial films was higher than their polycrystalline counterparts (~0.0002%), under illumination from an ultra violet light source [19]. Since the field of ferroelectric photovoltaic is young and evolving, reports on photovoltaic response of lead based thin films for other compositions are just evolving. For this work, thin films of composition $Pb_{0.95}La_{0.05}Zr_{0.54}Ti_{0.46}O_3$ (5/54/46) was chosen, which is reported to show maximum photo voltage in the ceramic form.

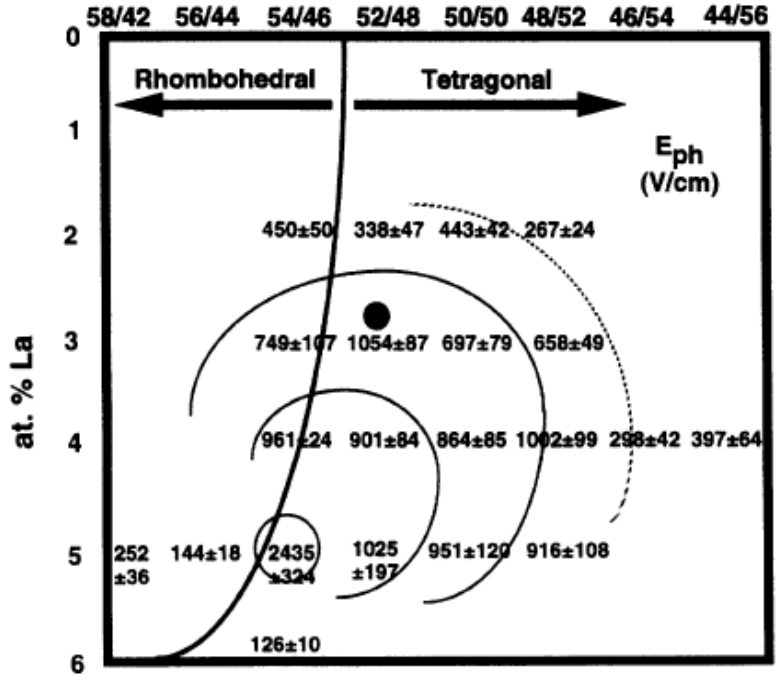

**Figure 1.5** Contour map of photovoltaic response in PLZT ceramics. Photo voltage is found to be maximum at 5/54/46 [18].

Improving the texture of the film is one among many strategies used to increase the photo-voltaic efficiency. Epitaxial films grown using single crystal substrates are reported to have higher photo voltaic efficiencies than their corresponding polycrystalline form [16,19,23].

Modification of the internal electric field of the ferroelectric is another strategy used to improve the photo response which is investigated in this work. The short circuit current ($I_{sc}$) generated in ferroelectrics, arises from the internal electric field induced by the ferroelectric polarization. This internal field, is composed of two components,

(i) the depolarization field ($E_{dp}$) originating from the polarization alignment in the bulk of the film, and
(ii) the interface electrical field ($E_i$) resulting from the electrode-ferroelectric interface.

As illustrated in figure 1.6, the depolarization field ($E_{dp}$) is polarization dependent, and is the expected component responsible for the photo response of the ferroelectric materials to electrical poling [19,20]. The interface electric field ($E_i$) is polarization independent, and it originates from the Schottky contacts of the electrode-ferroelectric interface [24-28].

Simulations done by Qin et al. on PLZT films show that PV efficiency can be improved by reducing the thickness of the film and higher photovoltaic efficiencies can be achieved using ultrathin ferroelectric films [20,29]. However, fabrication of ultrathin films with retention of ferroelectric property is a challenge.

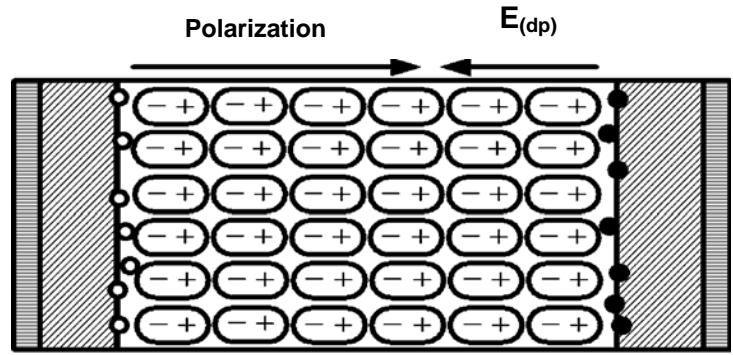

(a)

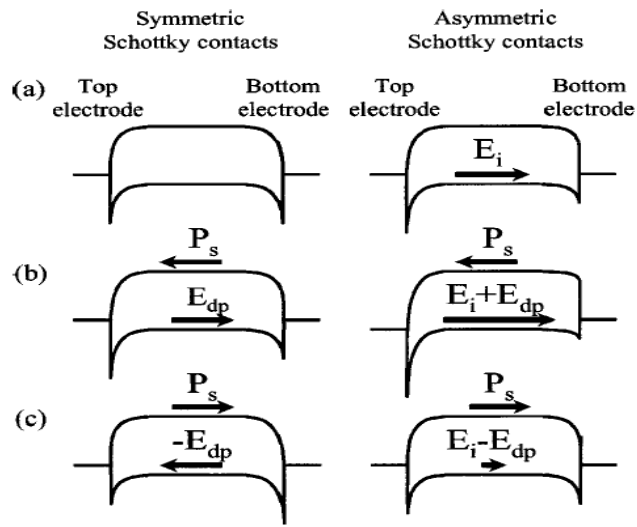

(b)

**Figure 1.6 (**a) The interfacial effect on the photo voltage in a ferroelectric material showing depolarized field ($E_{dp}$) [20]. (b) Diagram of metal-ferroelectric-metal capacitors having symmetric and asymmetric Schottky contacts indicating the interface electric field $E_i$ [28].

## 1.4 Organization of the Dissertation

This dissertation focuses on investigating ferroelectric $Pb_{0.95}La_{0.05}Zr_{0.54}Ti_{0.46}O_3$ (PLZT) films for photovoltaic applications. The dissertation is written in article style format and includes seven chapters which are based on five journal articles, and each chapter has its own bibliography.

Chapter 2 describes the preparation of $Pb_{0.95}La_{0.05}Zr_{0.54}Ti_{0.46}O_3$ thin films using solution based (MOD) method. The details on the fabrication process, optimization of structural and ferroelectric properties with process parameters are discussed. This article was published in the Journal of Integrated Ferroelectrics [30].

Chapter 3 discusses photovoltaic responses measured on post annealed $Pb_{0.95}La_{0.05}Zr_{0.54}Ti_{0.46}O_3$ thin films. The post deposition annealing was found to influence the open circuit voltage $V_{oc}$ and short circuit current $I_{sc}$ of the capacitor type devices. The influence of illumination on the ferroelectric properties is discussed in this chapter. This article has been submitted to Journal of Ferroelectrics.

Chapter 4 discusses about improving photovoltaic responses and the efficiency of $Pb_{0.95}La_{0.05}Zr_{0.54}Ti_{0.46}O_3$ thin film devices, by choosing a low work function metal (Al) as the top electrode. This article was published in Applied Physics Letters [31].

Chapter 5 describes about modeling the current–voltage (I-V) characteristics of $Pb_{0.95}La_{0.05}Zr_{0.54}Ti_{0.46}O_3$ films as a function of irradiance. An electric circuit model was developed and behavior of single cell photovoltaic device under various irradiance levels was predicted. This article was published in International Review on Modeling and Simulations [32].

Chapter 6 describes growth and characterization of Indium Tin Oxide (ITO) using pulsed electron deposition technique. The details on the deposition technique, optimization of optical and electrical properties of deposited ITO thin films with process parameters are discussed. This article was published in Journal of Vacuum Science and Technology A [33].

Chapter 7 presents the conclusion of the dissertation which summarizes the results of the overall research project. This chapter provides a brief summary of findings from all the five research articles. Suggestions for pursuing future research are provided in this chapter.

## 1.5 Chapter 1 References

1   F. F. Lange, "Chemical solution routes to single-crystal thin films," Science, 273 (5277), (1996), p. 903-909.

2   J. Ricote, S. Holgado, Z. Huang, P. Ramos, R. Fernandez, and M. L. Calzada, "Fabrication of continuous ultrathin ferroelectric films by chemical solution deposition methods," Journal of Materials Research, 23 (10), (2008), p. 2787-2795.

3   J. H. Kim and F. F. Lange, "Epitaxial growth of $PbZr_{0.5}Ti_{0.5}O_3$ thin films on (001) $LaAlO_3$ by the chemical solution deposition method," Journal of Materials Research, 14 (10), (1999), p. 4004-4010.

4   N.R.E.L, in *http://www.nrel.gov/ncpv/images/efficiency_chart.jpg* (NREL Website, 2011).

5   Ferroelectric Devices, K. Uchino, CRC Press, 2000

6   H. Han, S. Kotru, H. Zhong, and R. K. Pandey, "Effect of Nb doping on pyroelectric property of lead zirconate titanate films prepared by chemical solution deposition," Infrared Physics & Technology, 51 (3), (2008), p. 216-220.

7   O. Auciello, J. F. Scott, and R. Ramesh, "The Physics of Ferroelectric Memories," Physics Today, 51 (7), (1998), p. 22-27.

8   "An american national standard IEEE standard definitions of terms associated with ferroelectric and related materials," Ultrasonics, Ferroelectrics and Frequency Control, IEEE Transactions on, 50 (12), (2003), p. 1-32.

9   K. Yao, in *http://spie.org/x40422.xml* (SPIE Website, 2010).

10  A. M. Glass, D. V. D. Linde, and T. J. Negran, "High voltage bulk photovoltaic effect and the photorefractive process in $LiNbO_3$," Applied Physics Letters, 25 (4), (1974), p. 233-235.

11  P.S. Brody, "Large polarization-dependent photovoltages in ceramic $BaTiO_3$ + 5 wt.% $CaTiO_3$," Solid State Communications, 12 (7), (1973), p. 673-676.

12  K. Nonaka, M. Akiyama, T. Hagio, and A. Takase, "Bulk Photovoltaic Effect in Reduced/Oxidized Lead Lanthanum Titanate Zirconate Ceramics," Japanese Journal of Applied Physics, 34, (1995), p. 2344.

13  Photoferroelectrics, V. M. Fridkin, Springer, 1979

14	M. F. Vladimir and B. N. Popov, "Anomalous photovoltaic effect in ferroelectrics," Soviet Physics Uspekhi, 21 (12), (1978), p. 981.

15	S. Y. Yang, L. W. Martin, S. J. Byrnes, T. E. Conry, S. R. Basu, D. Paran, L. Reichertz, J. Ihlefeld, C. Adamo, A. Melville, Y. H. Chu, C. H. Yang, J. L. Musfeldt, D. G. Schlom, J. W. Ager, Iii, and R. Ramesh, "Photovoltaic effects in $BiFeO_3$," Applied Physics Letters, 95 (6), (2009), p. 062909.

16	S. Y. Yang, J. Seidel, S. J. Byrnes, P. Shafer, C. H. Yang, M. D. Rossell, P. Yu, Y. H. Chu, J. F. Scott, J. W. Ager, L. W. Martin, and R. Ramesh, "Above-bandgap voltages from ferroelectric photovoltaic devices," Nature Nanotechnology, 5 (2), (2010), p. 143-147.

17	K. Uchino and M. Aizawa, "Photostrictive Actuator Using PLZT Ceramics," Japanese Journal of Applied Physics, 24S3 (1985), p. 139.

18	P. Poosanaas, K. Tonooka, I. R. Abothu, S. Komarneni, and K. Uchino, "Influence of Composition and Dopant on Photostriction in Lanthanum-Modified Lead Zirconate Titanate Ceramics," Journal of Intelligent Material Systems and Structures, 10 (6), (1999), p. 439-445.

19	Q. Meng, Y. Kui, and C. L. Yung, "Photovoltaic characteristics in polycrystalline and epitaxial $Pb_{0.97}La_{0.03}Zr_{0.52}Ti_{0.48}O_3$ ferroelectric thin films sandwiched between different top and bottom electrodes," Journal of Applied Physics, 105 (6), (2009), p. 061624.

20	Q. Meng, Y. Kui, and C. L. Yung, "High efficient photovoltaics in nanoscaled ferroelectric thin films," Applied Physics Letters, 93 (12), (2008), p. 122904.

21	M. Qin, K. Yao, Y. C. Liang, and B. K. Gan, "Stability and magnitude of photovoltage in ferroelectric $Pb_{0.97}La_{0.03}Zr_{0.52}Ti_{0.48}O_3$ thin films in multi-cycle UV light illumination," Integrated Ferroelectrics, 95, (2007), p. 105-116.

22	M. Qin, K. Yao, Y. C. Liang, and B. K. Gan, "Stability of photovoltage and trap of light-induced charges in ferroelectric $WO_3$-doped $Pb_{0.97}La_{0.03}Zr_{0.52}Ti_{0.48}O_3$ thin films," Applied Physics Letters, 91 (9), (2007), p. 092904.

23	W. Ji, K. Yao, and Y. C. Liang, "Bulk Photovoltaic Effect at Visible Wavelength in Epitaxial Ferroelectric $BiFeO_3$ Thin Films," Advanced Materials, 22 (15), (2010), p. 1763-1766.

24	A. Kholkin, O. Boiarkine, and N. Setter, "Transient photocurrents in lead zirconate titanate thin films," Applied Physics Letters, 72 (1), (1998), p. 130-132.

25  F. Zheng, J. Xu, L. Fang, M. Shen, and X. Wu, "Separation of the Schottky barrier and polarization effects on the photocurrent of Pt sandwiched Pb $Zr_{0.20}Ti_{0.80}O_3$ films," Applied Physics Letters, 93 (17), (2008), p. 172101-172103.

26  V. Yarmarkin, B. M. Goltsman, M. Kazanin, and V. Lemanov, "Barrier photovoltaic effects in PZT ferroelectric thin films," Physics of the Solid State, 42 (3), (2000), p. 522-527.

27  D. Cao, J. Xu, L. Fang, W. Dong, F. Zheng, and M. Shen, "Interface effect on the photocurrent: A comparative study on Pt sandwiched $Bi_{3.7}Nd_{0.3}Ti_3O_{12}$ and $PbZr_{0.2}Ti_{0.8}O_3$ films," Applied Physics Letters, 96 (19), (2010), p. 192101-192103.

28  Y. S. Yang, S. J. Lee, S. Yi, B. G. Chae, S. H. Lee, H. J. Joo, and M. S. Jang, "Schottky barrier effects in the photocurrent of sol--gel derived lead zirconate titanate thin film capacitors," Applied Physics Letters, 76 (6), (2000), p. 774-776.

29  M. Qin, K. Yao, Y. C. Liang, and S. Shannigrahi, "Thickness effects on photoinduced current in ferroelectric $Pb_{0.97}La_{0.03}Zr_{0.52}Ti_{0.48}O_3$ thin films," Journal of Applied Physics, 101 (1), (2007), p. 014104-014108.

30  V. N. Harshan and S. Kotru, "Effect of Annealing on Ferroelectric Properties of Lanthanum Modified Lead Zirconate Titanate Thin Films," Integrated Ferroelectrics, 130 (1), (2011), p. 73-83.

31  V. N. Harshan and S. Kotru, "Influence of work-function of top electrodes on the photovoltaic characteristics of $Pb_{0.95}La_{0.05}Zr_{0.54}Ti_{0.46}O_3$ thin film capacitors," Applied Physics Letters, 100 (17), (2012), p. 173901-173904.

32  Harshan V Nampoori, Y. Jiang, J. Abu Qahouq, and S. Kotru, "Irradiance Dependent Equivalent Model for PLZT based Ferroelectric Photovoltaic Devices," International Review on Modelling and Simulations, 5 (1), (2012),p. 517.

33  H. V. Nampoori, V. Rincon, M. Chen, and S. Kotru, "Evaluation of indium tin oxide films grown at room temperature by pulsed electron deposition," Journal of Vacuum Science & Technology A, 28, (2010), p. 671-674.

# CHAPTER 2

# EFFECT OF ANNEALING ON FERROELECTRIC PROPERTIES OF LANTHANUM MODIFIED LEAD ZIRCONATE TITANATE THIN FILMS [1]

**Abstract**

Thin films of lanthanum modified lead zirconate titanate with chemical formula ($Pb_{0.95}La_{0.05}Zr_{0.54}Ti_{0.46}O_3$) were prepared using spin coating method. The sol for the PLZT was prepared in house by metal organic decomposition (MOD) technique. The prepared films were annealed in the temperature range of 550-750°C in a rapid annealing furnace in flowing oxygen to promote crystallinity in the films. Effect of post deposition annealing temperatures on the structure/morphology and ferroelectric properties of the films were examined using X-ray diffraction, atomic force microscopy and ferroelectric tester. Parameters were optimized for obtaining films with excellent ferroelectric properties and low leakage currents. A high value of polarization ~ 72μC/cm$^2$ coupled with a minimum leakage current density of 10$^{-8}$A/cm$^2$ was obtained from films annealed at temperature of 750 °C. The results indicate that choice of proper annealing process is vital to control the structure and morphology which are important parameters to achieve good ferroelectric properties in PLZT films. These films are being used to study the photovoltaic response from such material for their potential use in ferroelectric photovoltaic devices.

**2.1  Introduction**

Lead zirconium titanate (PZT), a solid solution of $PbZrO_3$ and $PbTiO_3$, is the most extensively studied material system at the morphotropic phase boundary (MPB) composition with Zr/Ti ratio of 52/48. At the MPB composition, the tetragonal and rhombohedral phase co-exists resulting in very high values of ferroelectric and piezoelectric properties especially in the bulk. PZT system can be modified by doping with impurities to enhance various properties of the system. The effect of Nb doping on piezoelectric coefficient and texture of PZT films has been studied [1-4] to find the optimal Nb doping level leading to the maximum value of effective transverse piezoelectric coefficient -$e_{31,f}$ [3]

On the other hand, doping PZT with appropriate amount of La has been reported to enhance the optical properties of the system. Thus lanthanum modified PLZT system has become an equally important material. Due to the large electro-optic coefficient [5-8], this material system finds applications in optical MEMS, optical modulators/transducers and smart sensors [9-13]. In addition it is becoming a material of choice for other applications where combination of ferroelectric properties and the optical transparency are utilized to exploit the potential for future photovoltaic devices [14]. Recent reports on the use of ferroelectric materials towards non-conventional photovoltaic (PV) devices [15] due to their ability to exhibit above band gap voltages have generated interest for using the ferroelectric materials towards future PV devices [16]. Other ferroelectric oxides which have been studied for such applications include $BiFeO_3$ and $LiNbO_3$ [17,18]

The highest photovoltaic response in the PLZT ceramic system was observed in the 3/52/48 ($Pb_{0.97}La_{0.03}Zr_{0.52}Ti_{0.48}O_3$) composition by Uchino et al [13]. Later an extensive study on the photo response of various compositions of PLZT ceramics was done by Poosanaas et al [19] which shows that the maximum photo current and photo voltage are not obtained from the same composition. In their work, the maximum photocurrent was obtained for PLZT composition of 3/52/48 whereas the maximum photo voltage was obtained at PLZT composition of 5/54/46 ($Pb_{0.95}La_{0.05}Zr_{0.54}Ti_{0.46}O_3$). There is plethora of papers based on fabrication of PLZT films by various techniques including vacuum deposition processes such as PLD [20], sputtering [21] and electron-beam evaporation [22] as well as by solution based processing for various applications [10,23-25].However, all reported work on photo response of the PLZT films is based on $Pb_{0.97}La_{0.03}Zr_{0.52}Ti_{0.48}O_3$ (3/52/48) or other compositions [15,25-28].

In this work we have prepared PLZT films with composition $Pb_{0.95}La_{0.05}Zr_{0.54}Ti_{0.46}O_3$ (5/54/46), by MOD technique. The purpose of the present work is to investigate the effect of post deposition annealing temperature on the structural and electrical properties of $Pb_{0.95}La_{0.05}Zr_{0.54}Ti_{0.46}O_3$ thin films and to optimize processing conditions to achieve films with maximum polarization and low leakage current. Morphology/structure and ferroelectric properties of the films grown on platinized silicon substrates are reported here. These films are being used for investigating the photo response under illumination in UV and visible wavelength. To the best of our knowledge this is first attempt to fabricate PLZT thin films of 5/54/46 composition using solution based MOD or any other method.

## 2.2  Experimental Details

The composition of PLZT used in this study was 5/54/46 with formula $Pb_{0.95}La_{0.05}Zr_{0.54}Ti_{0.46}O_3$. The solution was prepared in-house using a MOD route reported by Han et al [29] for preparation of Nb doped PZT. The procedure was modified by doping lanthanum instead of Nb. The acetate salts of Pb and La were used and dissolved using 2-Methoxyethanol (2-M) as solvent. Zr-propoxide and Ti-butoxide were as added to the solution mixture, with intermediate heating and cooling cycles. The final solution was made to the desired concentration of 0.4 molar by adding 2-M solution. Figure 2.1 shows the typical process flow of our PLZT film deposition process.

Commercially available Pt/Si substrates purchased from Radiant Technologies were used in this work. All of the films were deposited using spin coating technique at room temperature, with a spin speed of 3000 rpm for 30 sec. Each layer was hydrolyzed at 120 °C for 2 minutes, followed by a pyrolysis step at 400 °C for 2 minutes to remove organics, using a hot plate. The process was repeated for three layers to achieve 210 nm thick films as measured by a profilometer (Dektak). At this point, all the three layers were annealed together, using a RTA furnace (Qualiflow) for 2 minutes, in flowing oxygen (2000 sccm). Four different temperatures viz., 550 °C, 650 °C, 700 °C and 750 °C were chosen to investigate the effect of post deposition annealing on various properties of the prepared films. An X-ray diffractometer with Cu Kα radiation at 40 kV (Rigaku) was used to study the structure of the PLZT films. The scanning rate was 1 °/min, the interval was 0.02 degree, and the scanning range was 20-60°. Roughness was estimated from data obtained from an atomic force microscope (Vecco). Circular platinum electrodes about 1000 Å thick were sputtered on the film using a shadow mask, with the area of electrodes being 9.1 x$10^{-8}$ m$^2$. These electrodes serve as contacts to measure the electrical properties. All electrical measurements were taken using symmetrical Pt electrodes in top-bottom capacitor configuration. To ensure proper electrical contact between two electrodes, the capacitances ($C_p$) was measured using a LCR (HP) meter before measuring the Polarization-Electric field (P-E) hysteresis loop using a probe station (Signatone) connected to Radiant Technology workstation. All hysteresis measurements were taken at a frequency of 1 kHz.

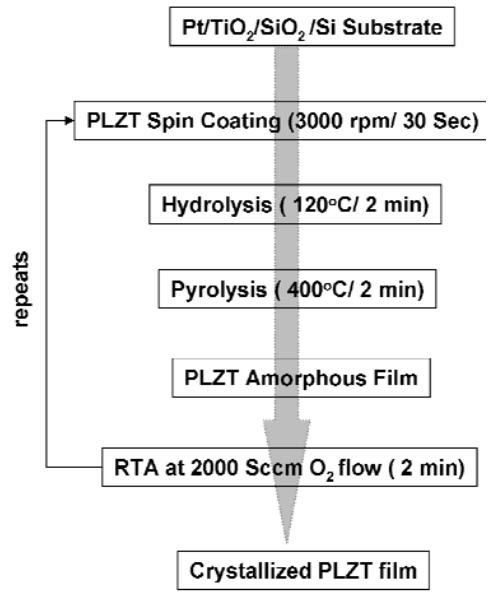

**Figure 2.1**. Process flow diagram for preparation of $Pb_{0.95}La_{0.05}Zr_{0.54}Ti_{0.46}O_3$ thin films.

### 2.3    Results and Discussion

Figure 2.2 shows the ($\theta$ -2$\theta$) scans from the X-ray diffraction patterns for the PLZT films annealed at various temperatures ranging from 500°C to 750°C. All the films are polycrystalline in nature, exhibiting only the perovskite (100) and (110) peaks with no presence of any pyrochlore phase. This result is similar to the earlier reported work by Qin et al. [28] of $Pb_{0.97}La_{0.03}Zr_{0.52}Ti_{0.48}O_3$ films on Pt substrates. The relative peak intensity of (110) and (100) is seen to increase with annealing temperature for all the films except for the film annealed at 650 °C. Figure 2.3 shows the surface morphology of each film as measured using AFM, and (RMS) roughness was found to be 4.74, 4.79, 6.35 and 7.61 nm for the films annealed at 550 °C, 650 °C, 700 °C and 750 °C respectively.

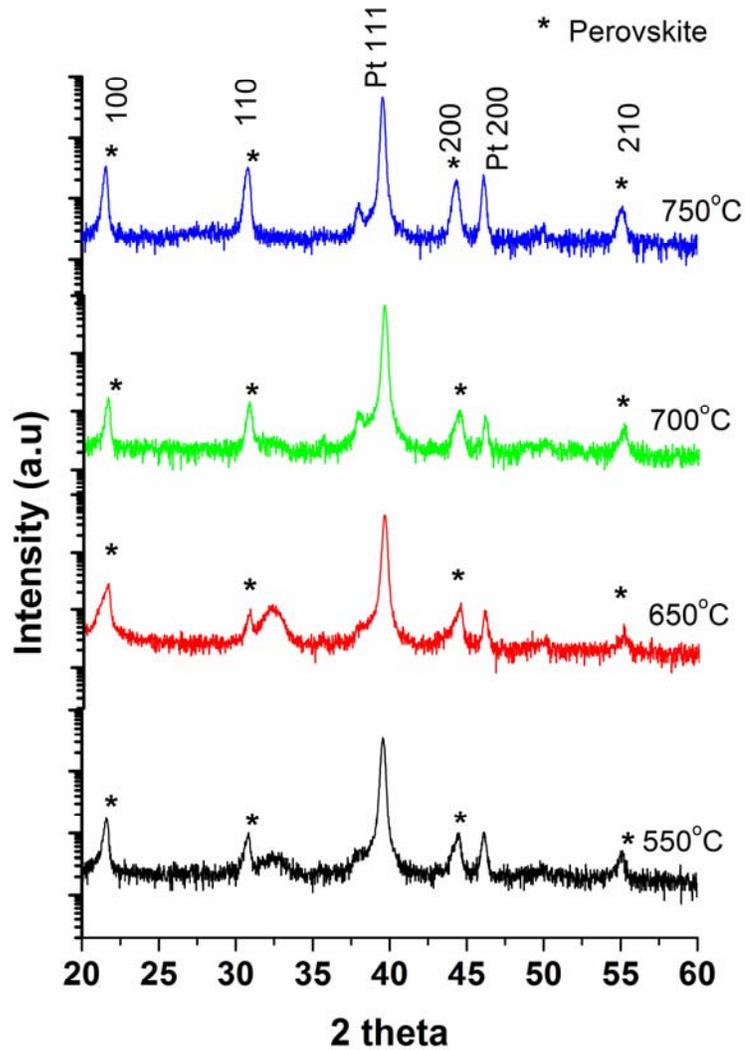

**Figure 2.2.** X-ray diffraction of $Pb_{0.95}La_{0.05}Zr_{0.54}Ti_{0.46}O_3$ films post annealed at 550 °C, 650 °C, 700 °C and 750 °C.

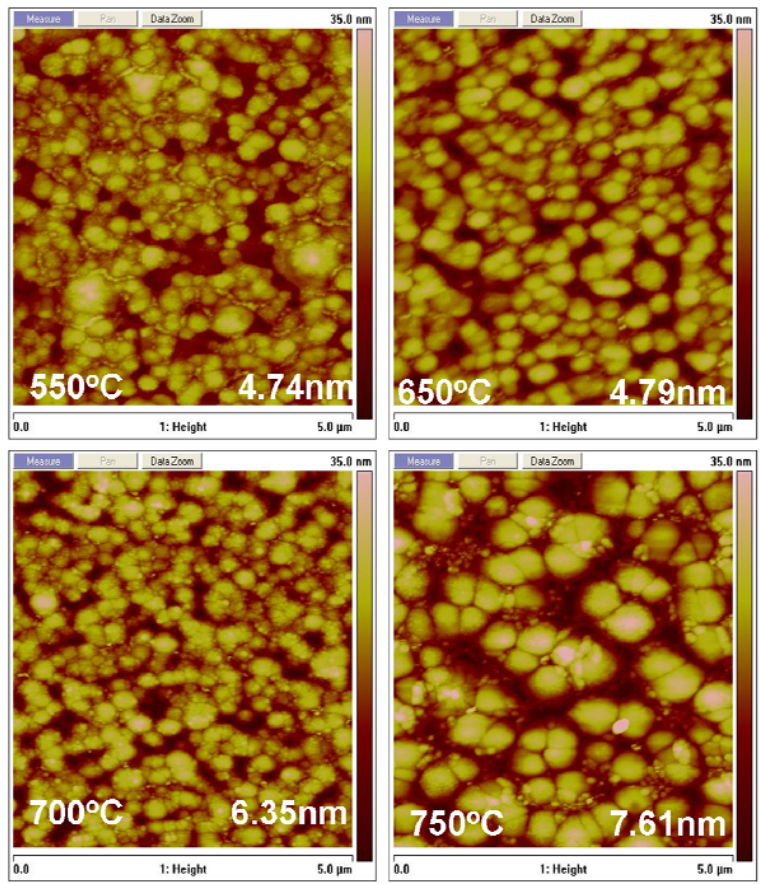

**Figure 2.3.** AFM images showing morphology of the as grown films, films were annealed at 550 °C, 650 °C, 700 °C and 750 °C.

The dielectric constant $\varepsilon_r$ as a function of bias voltage is shown in figure 2.4.1, and ferroelectric behaviors of the films measured as P-E curves is presented in figure 2.4.2. The measurements were taken at 1 kHz frequency at zero bias. The film annealed at 750 °C showed the highest value for dielectric constant. The value of dielectric constant dropped from 2075 for films annealed at 750 °C to 1279 for films annealed at 550 °C.

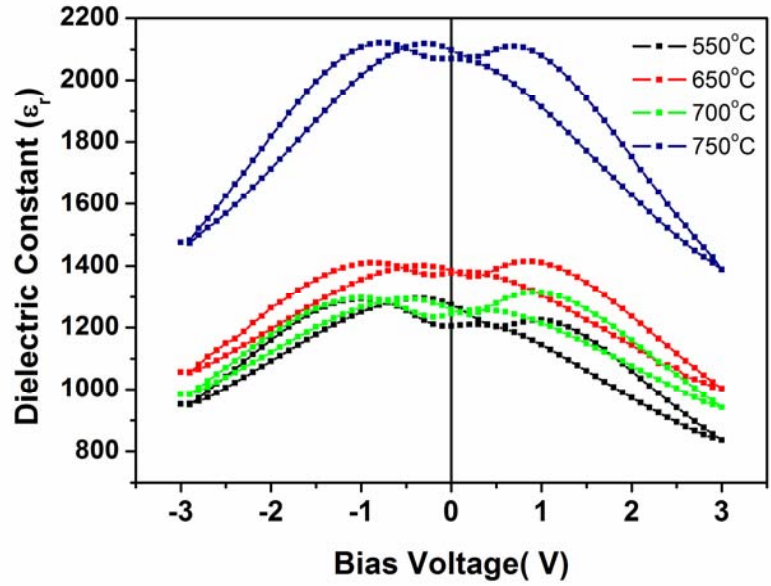

**Figure 2.4.1** Dielectric Constant ($\varepsilon_r$) vs bias voltage for Pb$_{0.95}$La$_{0.05}$Zr$_{0.54}$Ti$_{0.46}$O$_3$ (PLZT) thin films annealed in temperature range of 550-750$^{\circ}$C.

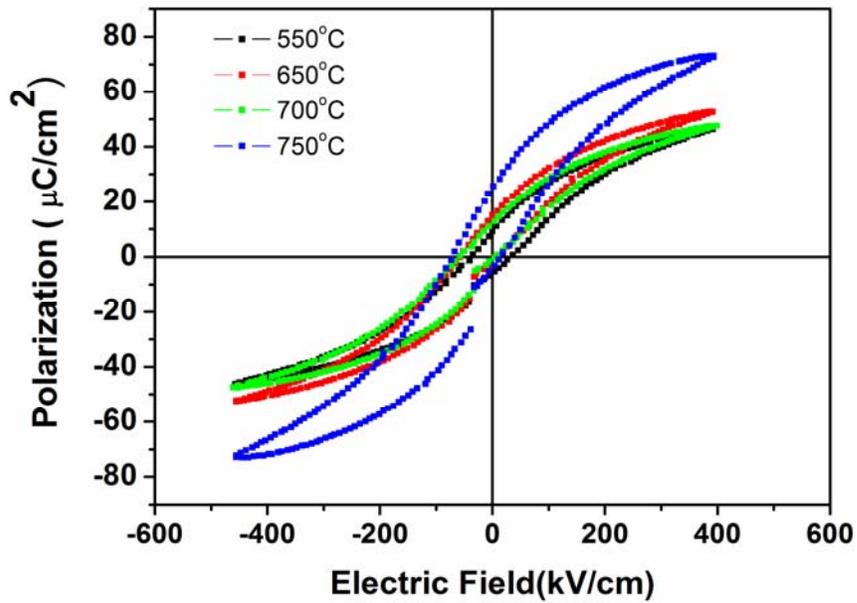

**Figure 2.4.2** The Polarization vs Electric field of 210 nm Pb$_{0.95}$La$_{0.05}$Zr$_{0.54}$Ti$_{0.46}$O$_3$ films post annealed at various temperatures.

All the samples show ferroelectric switching loop behavior indicating good ferroelectric properties. Films annealed at the 750 °C were seen to have the saturation polarization ($P_s$) of 72 µC/cm$^2$ with a remnant polarization ($P_r$) of 23 µC/cm$^2$ and an approximate coercive field ($E_c$) of 25 kV/cm. The polarization values were seen to decrease for films annealed at lower temperatures. The $P_s$ values obtained from our films are higher than reported values for Pb$_{0.97}$La$_{0.03}$Zr$_{0.52}$Ti$_{0.48}$O$_3$ films grown by solution process with similar thickness (196 nm), but for a different composition [28].

Figure 2.5 shows leakage current characteristics (I-V) of the film as a function of applied electrical field. From figure 2.5, it is clear that current-voltage behavior for all the films annealed at various temperatures is almost symmetric in both quadrants. The sample grown at 750 °C which showed the maximum polarization was also seen to exhibit the lowest leakage current of the order 10$^{-8}$ A/cm$^2$.

The origin of reduction of leakage current density in films annealed at higher temperatures is attributed to the change of film morphology as observed from AFM results from Figure 2.3. Films annealed at the highest temperature are seen to have the highest roughness of 7.61 nm with larger grain size. The roughness was seen to decrease from 7.61 to 4.74 nm as the annealing temperature was reduced from 750 to 550 °C. However, the grain size is also seen to decrease with decrease in post deposition temperature.

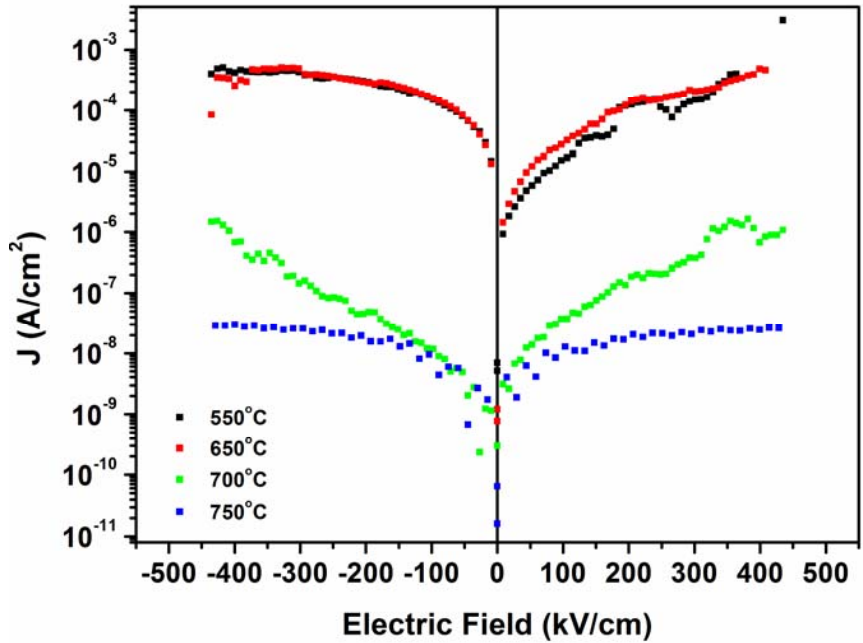

**Figure 2.5.** Leakage current as a function of electric field for $Pb_{0.95}La_{0.05}Zr_{0.54}Ti_{0.46}O_3$ thin films annealed at 550-750°C.

In general, grain boundaries are known to be a path for leakage of current. In our films since the electrical properties are measured with top-bottom configuration, the observed reduction of leakage current and increase in polarization could be attributed to the increase in grain size locally. Similar trends of lowering leakage current with increase in grain sizes have been reported for a variety of PLZT films [30-32]. Results obtained for Saturation polarization ($P_s$), Remanant Polarization ($P_r$), Coercive Field ($E_c$), leakage current density ($J/cm^2$), dielectric constant ($\varepsilon_r$), obtained at four annealing temperatures are compiled in Table 2.1

**Table 2.1:** Saturation Polarization ($P_s$), Remnant Polarization ($P_r$), Coercive Field ($E_c$), Leakage current density ($J/cm^2$), Dielectric constant ($\varepsilon_r$) for PLZT ($Pb_{0.95}La_{0.05}Zr_{0.54}Ti_{0.46}O_3$) thin films prepared by spin coating and post annealed at various temperatures using rapid thermal annealing process.

| Post deposition annealing temperature (°C) | $P_s$ ($\mu C/cm^2$) | $P_r$ ($\mu C/cm^2$) | $E_c$ (kV/cm) | Leakage current $J(A/cm^2)$ (at 0V) | Dielectric Constant ($\varepsilon_r$) |
|---|---|---|---|---|---|
| 550 | 46.49 | 10.56 | 31.46 | $4.87 \times 10^{-09}$ | 1279 |
| 650 | 52.78 | 14.92 | 28.29 | $7.92 \times 10^{-09}$ | 1270 |
| 700 | 46.49 | 14.56 | 25.15 | $3.01 \times 10^{-10}$ | 1385 |
| 750 | 72.40 | 23.77 | 22.01 | $1.52 \times 10^{-11}$ | 2075 |

Leakage current mechanisms in lead based thin film capacitors are predominantly found to have either interface limited (Schottky emission) or bulk limited (Poole Frenkel (PF) emission) [20,33,34]. The Schottky effect is very similar to the thermionic emission, where the carriers are emitted from the electrode surface to the conduction band of the semiconductor (PLZT), which is aided, due to the applied field and the potential barrier (Φ) height. For a Schottky type leakage emission the current-voltage characteristics are given by

$$J = A^*T^2 \frac{(e^{-(\phi-\beta_s V^{1/2})})}{KT} \quad \text{where} \tag{1}$$

$$\beta_s = \left\{ \frac{q^3}{(4\pi\varepsilon_\circ Kd)^{\frac{1}{2}}} \right\}$$

Here, A* is the modified Richardson constant, Φ is the interfacial barrier height and $\beta_s V^{1/2}$ is the lowering of barrier height due to image force interaction and applied field [20]. Poole Frenkel (PF) emission is the field assisted thermal ionization of trapped carries into the conduction band of the PLZT and the PF conductivity is given by

$$\sigma = \frac{(ce^{-(E_1 - \beta_{PF} V^{1/2})})}{kT} \quad \text{where} \tag{2}$$

$$\beta_{PF} = \left( \frac{q^3}{\pi\varepsilon_o Kd} \right)^{\frac{1}{2}}$$

$E_1$ is the trap ionization energy and $\beta_{PF} V^{1/2}$ is the barrier lowering due to the applied field.

To understand the controlling mechanism of leakage conduction of our PLZT thin films, we analyzed the leakage data of the films with respect to both Schottky and Poole Frenkel models. Using the I-V data (presented in Figure 2.5) and equations (1) and (2), log (J/T$^2$) and (log σ) for all the films annealed at four different temperatures were computed. The log (J/T$^2$) vs V$^{1/2}$ (Schottky type) and log σ vs V$^{1/2}$ (PF type) were plotted for each film. Figure 6 shows log (J/T$^2$) vs V$^{1/2}$ obtained from the current in positive quadrant of I-V curve for all the films. It can be observed that log (J/T$^2$) vs V$^{1/2}$ plots follow a straight line, which is typical for Schottky conduction behavior. Such a linear behavior was not observed in case of the log σ vs V$^{1/2}$ of the IV data (not shown). The linear behavior seen from figure 2.6, indicates that the Schottky leakage mechanism is predominant and favorable in our PLZT films.

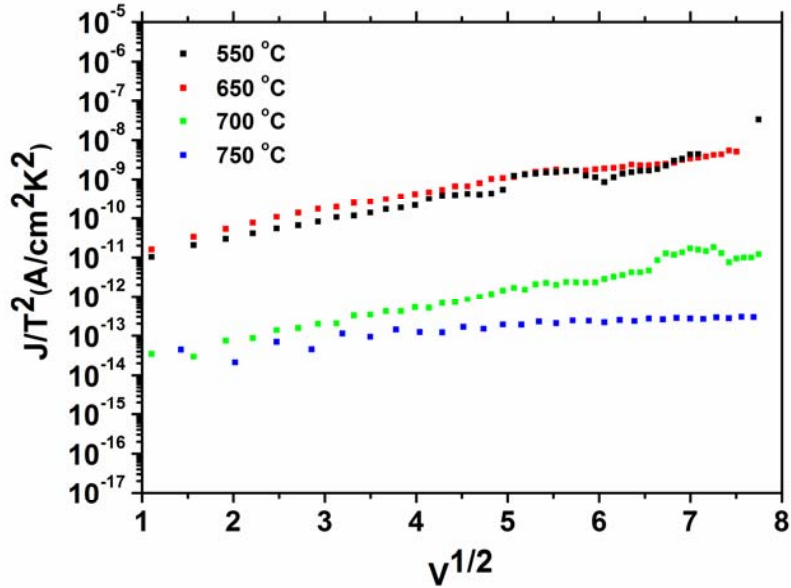

**Figure 2.6.** Plot of log $(J/T^2)$ vs $V^{1/2}$ obtained from the current in positive quadrant of I-V curve of figure 2.5; the trend suggests that the PLZT films follow Schottky conduction mechanism.

**2.4    Conclusion**

$Pb_{0.95}La_{0.05}Zr_{0.54}Ti_{0.46}O_3$ thin films were prepared by spin coating method using a MOD prepared sol. Films were annealed after deposition in the temperature range of 550-750°C in order to optimize the structural and ferroelectric properties. The roughness and grain size of the films was seen to depend on the post deposition annealing temperatures. Films annealed at 750$^{\circ}$C showed the highest values for saturation polarization (72 μC/cm$^2$) as well as minimum leakage current (10$^{-8}$A/cm$^2$). The higher values of dielectric constant and saturation polarization coupled with the lowest leakage current clearly indicate that the post deposition annealing temperature of 750$^{\circ}$C is the optimum temperature to achieve films with best ferroelectric properties for the PLZT (5/54/46) composition. The results indicate that the structural and morphological properties play a vital role on the ferroelectric properties of these films and special consideration should be given to post deposition annealing temperatures to control the structural and ferroelectric properties. Schottky leakage mechanism was seen to dominate in these PLZT films.

## 2.5    Acknowledgments

This work was supported by the National Science Foundation under ECCS Grant No. 0943711. Authors extend thanks to the Center for Materials and Information Technology (MINT), University of Alabama, for providing the X-ray and AFM facilities. The work was also partially supported from the startup funds provided by ECE, College of Engineering, UA.

## 2.6 Chapter 2 References

1    T. Haccart, E. Cattan, D. Remiens, S. Hiboux, and P. Muralt, "Evaluation of niobium effects on the longitudinal piezoelectric coefficients of Pb(Zr, Ti)$O_3$ thin films," Applied Physics Letters, 76 (22), (2000), p. 3292-3294.

2    K. W. Kwok, R. C. W. Tsang, H. L. W. Chan, and C. L. Choy, Effects of niobium doping on the piezoelectric properties of sol-gel-derived lead-zirconate titanate films, J. Appl. Phys 95 (3), (2004), p. 1372-1376.

3    Jian Zhong, Sushma Kotru, Hui Han, John Jackson, and R. K. Pandey, "Effect of Nb Doping on Highly {100}-textured PZT Films Grown on CSD-prepared PbTi$O_3$ Seed Layers," Integrated Ferroelectrics, (Accepted for print), (2011).

4    K. K. P. Kwok. K. W, Tsang. R.C.W, Chan. H. L. W, Choy. C. L. , "Preparation and piezoelectric properties of sol-gel-derived Nb-doped PZT films for MEMS applications," Integrated Ferroelectrics, 80, (2006), p. 107.

5    G. H. Haertling, "Ferroelectric Ceramics: History and Technology," Journal of the American Ceramic Society, 82 (4), (1999), p. 797-818.

6    Y. Xi, C. Zhili, and L. E. Cross, Polarization and depolarization behavior of hot pressed lead lanthanum zirconate titanate ceramics, Journal of Applied Physics, **54** (6), (1983), p. 3399-3403.

7    N. Setter, D. Damjanovic, L. Eng, G. Fox, S. Gevorgian, S. Hong, A. Kingon, H. Kohlstedt, N. Y. Park, G. B. Stephenson, I. Stolitchnov, A. K. Taganstev, D. V. Taylor, T. Yamada, and S. Streiffer, "Ferroelectric thin films: Review of materials, properties, and applications," Journal of Applied Physics, 100 (5), (2006), p. 46.

8    D. Dimos, W. L. Warren, M. B. Sinclair, B. A. Tuttle, and R. W. Schwartz, "Photoinduced hysteresis changes and optical storage in (Pb,La)(Zr,Ti)$O_3$ thin films and ceramics," Journal of Applied Physics, 76 (7), (1994), p. 4305-4315.

9    M. Ichiki, Y. Morikawa, and T. Nakada, "Electrical Properties of Ferroelectric Lead Lanthanum Zirconate Titanate as an Energy Transducer for Application to Electrostatic-Optical Motor," Japanese Journal of Applied Physics, 41 (11B), (2002),p. 6993.

10    K. Uchiyama, A. Kasamatsu, Y. Otani, and T. Shiosaki, "Electro-Optic Properties of Lanthanum-Modified Lead Zirconate Titanate Thin Films Epitaxially Grown by the Advanced Sol Gel Method," Japanese Journal of Applied Physics, 46, (2007), p. L244.

11   H. Shima, T. Iijima, H. Funakubo, T. Nakajima, H. Naganuma, and S. Okamura, "Electrooptic and Piezoelectric Properties of (Pb,La)(Zr,Ti)$O_3$ Films with Various Zr/Ti Ratios," Japanese Journal of Applied Physics, 47, (9) ,(2008) p. 7541.

12   A. Zomorrodian, N. J. Wu, Y. Song, S. Stahl, A. Ignatiev, E. B. Trexler, and C. A. Garcia, "Micro Photo Detector Fabricated of Ferroelectric;Metal Heterostructure," Japanese Journal of Applied Physics, 44 ,(2005), p. 6105.

13   K. Uchino and M. Aizawa, "Photostrictive Actuator Using PLZT Ceramics," Japanese Journal of Applied Physics, 24, (S3),(1985) p. 139.

14   H. T. Huang, "SOLAR ENERGY Ferroelectric photovoltaics," Nature Photonics, 4 (3), (2010), p. 134-135.

15   M. Qin, K. Yao, and Y. C. Liang, "High efficient photovoltaics in nanoscaled ferroelectric thin films," Applied Physics Letters, 93 (12), (2008),p. 122904.

16   S. Y. Yang, J. Seidel, S. J. Byrnes, P. Shafer, C. H. Yang, M. D. Rossell, P. Yu, Y. H. Chu, J. F. Scott, J. W. Ager, L. W. Martin, and R. Ramesh, "Above-bandgap voltages from ferroelectric photovoltaic devices," Nature Nanotechnology, 5 (2) p. 143-147.

17   S. Y. Yang, L. W. Martin, S. J. Byrnes, T. E. Conry, S. R. Basu, D. Paran, L. Reichertz, J. Ihlefeld, C. Adamo, A. Melville, Y. H. Chu, C. H. Yang, J. L. Musfeldt, D. G. Schlom, J. W. Ager, and R. Ramesh, "Photovoltaic effects in BiFe$O_3$," Applied Physics Letters, 95 (6), (2009), p.143

18   A. M. Glass, D. V. D. Linde, and T. J. Negran, "High voltage bulk photovoltaic effect and the photorefractive process in LiNb$O_3$", Applied Physics Letters, 25 (4), (1974), p. 233-235.

19   P. Poosanaas, K. Tonooka, I. R. Abothu, S. Komarneni, and K. Uchino, "Influence of Composition and Dopant on Photostriction in Lanthanum-Modified Lead Zirconate Titanate Ceramics," Journal of Intelligent Material Systems and Structures, 10 (6), (1999), p. 439-445.

20   B. Nagaraj, S. Aggarwal, T. K. Song, T. Sawhney, and R. Ramesh, "Leakage current mechanisms in lead-based thin-film ferroelectric capacitors," Physical Review B, 59 (24), (1999), p. 16022.

21   M. Okuyama, T. Usuki, Y. Hamakawa, and T. Nakagawa, "Epitaxial growth of ferroelectric PLZT thin film and their optical properties," Applied Physics A: Materials Science & Processing, 21 (4), (1980), p. 339-343.

22   B. Panda, S. K. Ray, A. Dhar, A. Sarkar, D. Bhattacharya, and K. L. Chopra, "Electron beam deposited lead lanthanum zirconate titanate thin films for silicon based device applications", Journal of Applied. Phys. **79**, (1996) p. 1008.

23  C. D. E. Lakeman and D. A. Payne, "Sol-gel processing of electrical and magnetic ceramics," Materials Chemistry and Physics, 38 (4), (1994), p. 305-324.

24  M. Sayer, G. Yi, and M. Sedlar, "Comparative Sol gel processing of PZT thin films," Integrated Ferroelectrics, 7 (1-4), (1995), p. 247-258.

25  D. Dimos, B. G. Potter, M. B. Sinclair, B. A. Tuttle, and W. L. Warren, "Photo-induced and electrooptic properties of (Pb,La)(Zr,Ti)$O_3$ films for optical memories," Integrated Ferroelectrics, 5 (1), (1994), p. 47-58.

26  M. Qin, K. Yao, Y. C. Liang, and B. K. Gan, "Stability of photovoltage and trap of light-induced charges in ferroelectric $WO_3$-doped $(Pb_{0.97}La_{0.03})(Zr_{0.52}Ti_{0.48})O_3$ thin films," Applied Physics Letters, 91 (9), (2007), p. 092904.

27  M. Qin, K. Yao, and Y. C. Liang, "Photovoltaic mechanisms in ferroelectric thin films with the effects of the electrodes and interfaces," Applied Physics Letters, 95 (2), (2009), p. 022912-022913.

28  M. Qin, K. Yao, and Y. C. Liang, "Photovoltaic characteristics in polycrystalline and epitaxial $(Pb_{0.97}La_{0.03})(Zr_{0.52}Ti_{0.48})O_3$ ferroelectric thin films sandwiched between different top and bottom electrodes," Journal of Applied Physics, 105 (6), (2009), p.61624.

29  H. Han, X. Y. Song, J. Zhong, S. Kotru, P. Padmini, and R. K. Pandey, "Highly a-axis-oriented Nb-doped $Pb(Ti_xZr_{1-x})O_3$ thin films grown by sol-gel technique for uncooled infrared dectors," Applied Physics Letters, 85 (22), (2004), p. 5310-5312.

30  J.-E. Y. In-Seok Lee, Dong-Hyun Hwang, Sang-Jih Kim, Jung-Hoon Ahn, Young Guk Son," The ferroelectric and electrical properties of PLZT thin films" in Journal of Ceramic Processing Research (Korea, 2009), Vol. 10, pp. 541-543.

31  S. M. Cho and D. Y. Jeon, "Effect of annealing conditions on the leakage current characteristics of ferroelectric PZT thin films grown by sol-gel process," Thin Solid Films, 338 (1-2), (1999), p. 149-154.

32  M. Es-Souni, M. Abed, A. Piorra, S. Malinowski, and V. Zaporojtchenko, "Microstructure and properties of sol-gel processed $Pb_{1-x}La_x(Zr_{0.52},Ti_{0.48})_{1-x/4}O_3$ thin films. The effects of lanthanum content and bottom electrodes," Thin Solid Films, 389 (1-2), (2001), p. 99-107.

33  X. Chen, A. I. Kingon, H. Al-Shreef, and K. R. Bellur, "Electrical transport and dielectric breakdown in $Pb(Zr,Ti)O_3$ thin films," Ferroelectrics, 151 (1), (1994), p. 133-138.

34    J. F. Scott, B. M. Melnick, C. A. Araujo, L. D. Mcmillan, and R. Zuleeg, "D.C. leakage currents in ferroelectric memories," Integrated Ferroelectrics, 1 (2-4), (1992), p. 323-331.

# CHAPTER 3

# PHOTOVOLTAIC AND FERROELECTRIC PROPERTIES OF $Pb_{0.95}La_{0.05}Zr_{0.54}Ti_{0.46}O_3$ THIN FILMS UNDER DARK AND ILLUMINATED CONDITIONS [2]

**Abstract**

Thin films of $Pb_{0.95}La_{0.05}Zr_{0.54}Ti_{0.46}O_3$ were prepared using chemical solution method. The photovoltaic response of the films under dark and illuminated conditions was investigated using capacitor type devices with Pt as top and bottom electrodes. The results show that the open circuit voltage and the short circuit current of a capacitor type cell can be manipulated by controlling the post deposition annealing temperature of the films. Films annealed at 750˚C were seen to exhibit the highest short circuit current of 15.83 nA. The influence of illumination on properties such as capacitance, ferroelectric polarization and leakage current was studied. The capacitance of the film was seen to decrease when illuminated with light (60mW/cm$^2$), suggesting that illumination leads to generation of charge carriers. The films became leaky under illumination, resulting in an increase in leakage current by an order of magnitude compared to the dark conditions. The saturation polarization ($P_s$) was suppressed, whereas the remnant polarization ($P_r$) was not affected under illumination. Further, a shift in the hysteresis loop, along the voltage axis, was observed under illuminated conditions. This shift is attributed to the change of a space-charge field with illumination.

---

[2] This work is submitted as **"Photovoltaic And Ferroelectric Properties Of $Pb_{0.95}La_{0.05}Zr_{.54}Ti_{0.46}O_3$ Thin Films Under Dark And Illuminated Conditions"** by Harshan V N and Sushma Kotru to Journal of Ferroelectrics

## 3.1 Introduction

Ferroelectric materials have become increasingly important due to their multifunctional properties. Ferroelectric thin films are used in non volatile memory, high frequency electronics as well as micro electronics applications [1, 2]. La doped PZT, in additional to being a ferroelectric, is also optically transparent, thereby making it a material of choice for optical applications such as optical MEMS [3-5]. Properties of PLZT are sensitive to the Zr/Ti ratio; thus, the choice of composition is dictated by the particular application [6].

Recently, La doped PZT has been studied for its use in photovoltaic (PV) applications [7-9]. One of the widely studied compositions for this application is $Pb_{0.97}La_{0.03}Zr_{0.52}Ti_{0.48}O_3$ (3/52/48). Several fabrication techniques are reported to grow thin films of this material, including PLD, sputtering, e-beam, and solution based methods [10-15].

Ferroelectric materials are known to have a built-in polarization field, which, among other factors, is greatly influenced by the structure, morphology, annealing temperature, time, and environment of the material. This built-in polarization field is responsible for the PV effect in ferroelectrics [16] and control of this field can be used to tune their photovoltaic properties [8, 17]. In addition to change in I-V characteristics, some other properties are also affected when the ferroelectric thin films are subject to illumination [18, 19]. Dimos et al. studied the photo-induced properties of solution based PZT and PLZT films, using various bottom electrodes and ITO top electrodes. The hysteresis loops from these films were seen to show a change in coercive voltage and a suppression of switchable polarization under UV, and near-UV, illumination [18, 20].

Kholkin et al. studied the influence of UV illumination on the piezoelectric properties of sol-gel prepared PZT films with Zr/Ti ratio of 45/55. In their work, films were poled in dark conditions, and under UV light. The piezoelectric coefficient of films was seen to increase when films were poled under illumination [21]. This effect was attributed to the trapping of electrons near the film-electrode interface. In an another work, Khokin et al. showed that illumination of PLZT films under UV light resulted in a decrease of piezoelectric co-efficient and dielectric permittivity [19].

Thin films of $Pb_{0.95}La_{0.05}Zr_{0.54}Ti_{0.46}O_3$ (5/54/46) were chosen for this study, based on our previous work with this composition [22]. The films were fabricated using solution based method [22, 23]. The effect of annealing, on the structure, morphology and ferroelectric properties, was reported earlier [22]. The purpose of this work was to study the photovoltaic and ferroelectric properties of these films under dark and illuminated conditions.

## 3.2    Experimental

To prepare the PLZT (5/54/46) films, "sol" was prepared in house using acetate salts of Pb and La and 2-Methoxyethanol (2-M) as solvent. Zr-propoxide and Ti-butoxide were added to the solution with intermediate heating and cooling. The final concentration of the solution was made 0.4 molar by adding 2-M solution. PLZT films of 210 nm thickness were prepared by spin coating on commercially available platinized silicon wafers [22]. The as-grown films were post annealed using an RTA (Quantiflow) with flowing oxygen of 2000 sccm for 2 minutes at temperatures of 550°C, 600°C, 700 °C and 750°C. Circular Platinum electrodes of 70 nm thickness and 9.1 x$10^{-8}$ m$^2$ area were sputter deposited onto the films to serve as top electrodes for electrical measurements. The capacitance of the films was measured using an LCR meter (4284A from HP). A Precision Workstation (Radiant technologies) was used to measure ferroelectric properties and leakage currents. All measurements were done at room temperature. A setup (figure 3.1), built in-house, was used to measure the IV characteristics under dark and illuminated conditions. The sample was illuminated from the top using a Xe light source (SCI-Science Tech) and the light was focused onto the device using a concave mirror. The intensity of illumination was calibrated using a power meter (Newport 1918-C). All the measurements were carried out in top-bottom capacitor configuration, using a probe station (Signatone). The voltage bias was applied to the top electrode and current was measured using a source meter as shown in figure 3.1. A Keithely (6517) source meter, which was interfaced to a computer for data acquisition, was used to measure current (I) and voltage (V) data.

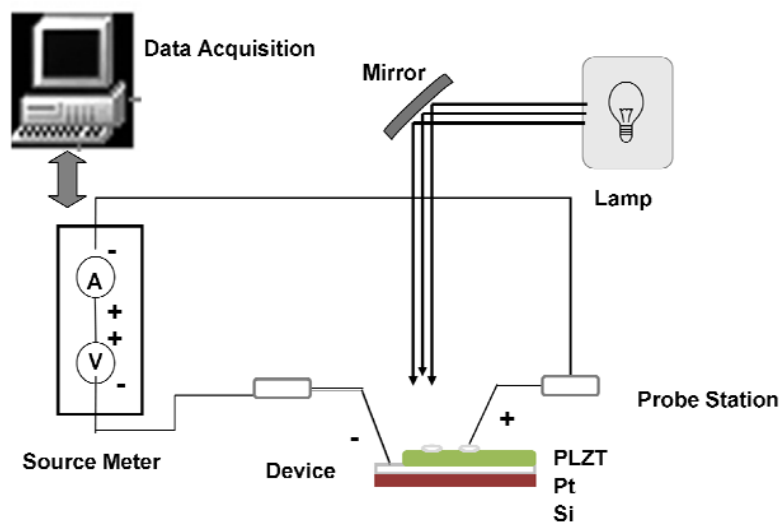

**Figure 3.1** A schematic of the setup used to measure the photovoltaic properties of PLZT capacitor type devices

### 3.3    Results and Discussion

Fig. 3.2 (a-d) shows the current-voltage (I-V) measurements of PLZT films annealed at 550°C, 650°C, 700°C and 750°C, when illuminated with light of intensity 60mW/cm$^2$. The current was measured in the voltage range of -3 V to +3 V. The maximum open circuit voltage ($V_{oc}$) was found to be -2.6 V for the film annealed at 550°C. $V_{oc}$ decreased from -2.56 V to -0.34 V as the annealing temperature of the films increased from 550°C to 750°C. A short circuit photocurrent ($I_{sc}$) of 4.20 nA was observed for the film annealed at 550°C and it increased to 15.83 nA for the sample annealed at 750°C. The $V_{oc}$ and $I_{sc}$ for these films are tabulated in Table 3.1.

The IV curves for the films are seen to have a positive y-intercept and a negative x-intercept. This trend is similar to that reported for ferroelectric films of BiFeO$_3$ [24] and La doped PZT's on oxide electrodes [7], but differs from our previous work [25]. The trend of IV curves of ferroelectric cells depends on the direction of the initial alignment of dipoles in the ferroelectric domains of the film.

To further understand the effect of illumination on ferroelectric properties of these films, the capacitance, polarization and leakage current of these films were measured under dark and illuminated conditions. For clarity, results from only one film sample (post annealed at 700°C) are included here.

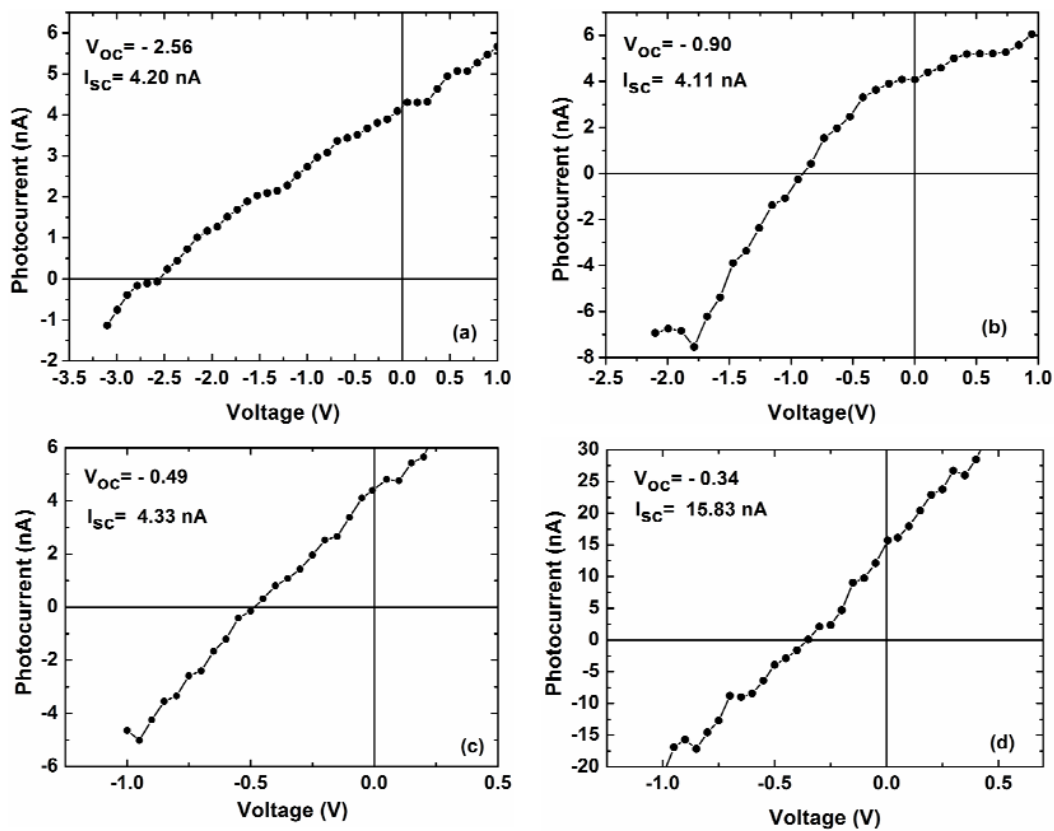

**Figure 3.2 (a,b,c,d)** The current-voltage curves of $Pb_{0.95}La_{0.05}Zr_{0.54}Ti_{0.46}O_3$ thin films post annealed at (a) 550 °C, (b) 650 °C, (c) 700°C and (d) 750 °C with Pt top electrodes.

Figure 3.3 represents the capacitance vs. voltage (CV) behavior of the film measured under dark and illuminated conditions. Both the CV curves show a butterfly behavior, which is typical of a ferroelectric material. The capacitance of the film is seen to decrease under illumination. The maximum change in capacitance of the films was observed at -1 V bias, which was found to decrease from 3.0 nF to 2.76 nF with illumination. The decrease in capacitance is an indication of generation of charge carriers upon illumination which results in increase in photo carrier density. A similar decrease in the c-axis dielectric constant was observed by Brody [26] in a barium titanate crystal in response to illumination.

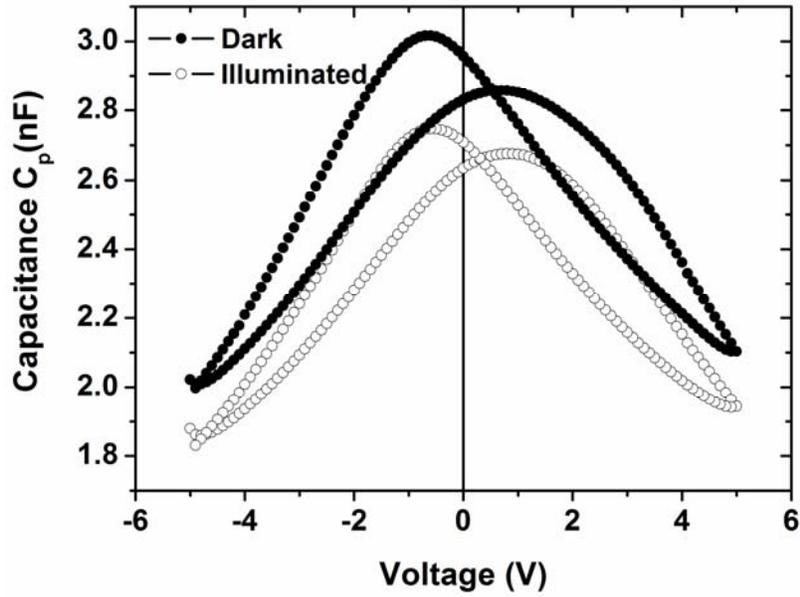

**Figure 3.3**: The capacitance vs. voltage behavior of the PLZT thin film post annealed at 700°C

Figure 3. 4 show the leakage current of the film measured in the range 10 V to -10 V, under dark and illuminated conditions. Under dark conditions, the leakage current is $10^{-8}$ A/cm$^2$ and it increases to $10^{-7}$ A/cm$^2$ when the film was illuminated. The leakage current is symmetrical in both quadrants. Such symmetrical increase in leakage current with illumination can be attributed to the generation of photo electrons in the film.

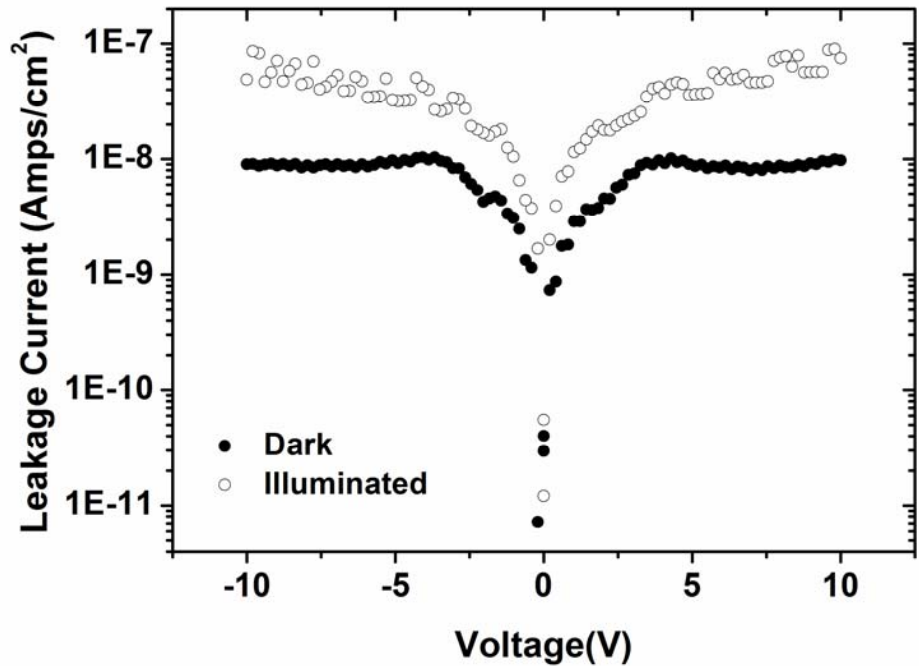

**Figure 3.4**: The leakage current vs. voltage behavior of the PLZT thin film annealed at 700$^{o}$C

The P-E characteristic of the film, measured under dark and illuminated conditions in the voltage range of -10 V to +10 V, are shown in figure 3.5. The maximum saturation polarization ($P_s$) is suppressed under illumination. The saturation polarization ($P_s$) under dark condition was ~ 42.45 µC/cm$^2$. The polarization ($P_s$) value was seen to reduce to ~ 33.69 µC/cm$^2$ and -37.08 µC/cm$^2$ in positive and negative direction respectively, under illumination. The remnant polarization (+$P_r$) was not affected under illumination. A shift in the hysteresis loop along the voltage axis was observed, which can be attributed to the charge carriers affecting the space-charge field with illumination.

A similar behavior in the ferroelectric hysteresis loop, under illumination, was reported by Kholkin et al. for PZT thin films prepared by sol-gel method [19]. They observed an increase in the squareness of the loops with illumination, which was attributed to compensation of the internal depolarization field of the ferroelectric due to generated photo induced charge carriers. Such a change in squareness, and shape of the hysteresis loops was not observed in our work. Polarization suppression in PZT films was also observed by Warren et al., in thin films of PZT and $BaTiO_3$ in [27].

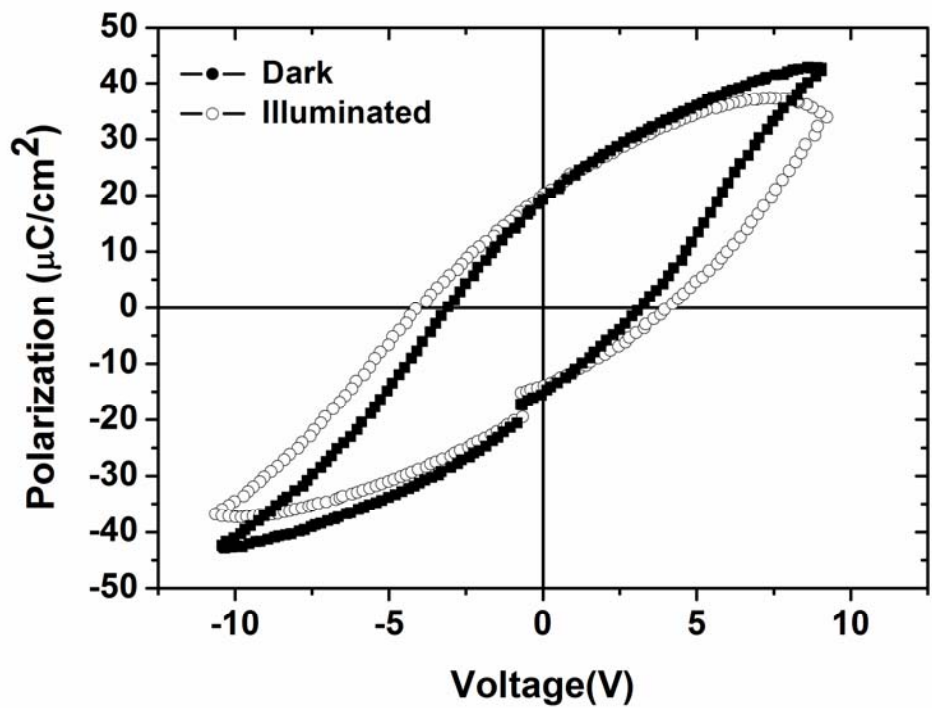

**Figure 3.5**: The P-E loops of PLZT thin film annealed at 700°C under dark and illuminated conditions

It is important to point out that the capacitance measured in dark shows an asymmetric CV curve, which could be due to an asymmetric depolarization field in the film.

Such an asymmetry is also evident from the polarization loops. Since both the top and the bottom electrode are similar (Pt), this asymmetric field can be attributed to the trapped charge carriers at the electrode interfaces arising from the difference of thermal cycling at the two electrodes. Similar results were reported by Kholkin [19, 28].

**Table 3.1:** The Photovoltaic characteristics for $Pb_{0.95}La_{0.05}Zr_{0.54}Ti_{0.46}O_3$ thin films.

| Annealing Temperature | Open Circuit Voltage ($V_{oc}$) | Short Circuit Current $I_{sc}$ (nA) |
|---|---|---|
| 550 °C | -2.56 | 04.20 |
| 650 °C | -0.90 | 4.11 |
| 700 °C | -0.49 | 4.33 |
| 750 | -0.34 | 15.83 |

### 3.4    Conclusion

PLZT films of 210 nm thickness were prepared by chemical solution and spin coating method. These films were annealed at different temperatures in order to optimize the ferroelectric properties. Capacitor type cells were prepared using Pt as top and bottom electrodes. The PV characteristics were measured under illumination. The $V_{oc}$ and $I_{sc}$ of these capacitor type cells were seen to change with the post annealing temperatures of PLZT films. The cells fabricated with PLZT films annealed at 750°C showed the highest short circuit current. The film also exhibited the highest polarization. This observation confirms that, to achieve optimum PV effect from ferroelectric cells, the polarization of the film needs to be optimized. The capacitance of the films was seen to decrease and leakage current was seen to increase with illumination, which is attributed to the generation of charge carriers in the film. The maximum saturation polarization ($P_s$) was seen to reduce and a shift in the hysteresis loop along the voltage axis was observed, under illumination.

## 3.5 Acknowledgements

The work was supported by the National Science Foundation under ECCS Grant No. 0943711 and the startup funds provided by the college of engineering, UA. Author (HVN) extends thanks to the electronic materials and device laboratory, University of Alabama, for the use of characterization facilities.

## 3.6 Chapter 3 References

1. Auciello O, Scott J F, Ramesh R, "The Physics of Ferroelectric Memories", Physics Today, 51, (1998), p. 22-27.

2. Setter N, Damjanovic D, Eng L, Fox G, Gevorgian S, Hong S, Kingon A, Kohlstedt H, Park N Y, Stephenson G B, Stolitchnov I, Taganstev A K, Taylor D V, Yamada T, Streiffer S, "Ferroelectric thin films: Review of materials, properties, and applications", J. Appl. Phys, 100, (2006), p. 051606–051652.

3. Zomorrodian A, Wu N J, Song Y, Stahl S, Ignatiev A, Trexler E B, Garcia C A, Micro Photo Detector Fabricated of Ferroelectric Metal Heterostructure. Jpn. J. Appl. Phys. 44, (2005), p. 6105-6108.

4. Ichiki M, Morikawa Y, Nakada T, "Electrical Properties of Ferroelectric Lead Lanthanum Zirconate Titanate as an Energy Transducer for Application to Electrostatic-Optical Motor", Jpn. J. Appl. Phys. 41, (2002), p. 6993-6996.

5. Uchino K, Aizawa M, "Photostrictive Actuator Using PLZT Ceramics", Jpn. J. Appl. Phys. 24S3, (1985), p. 139-141.

6. Poosanaas P, Tonooka K, Abothu I R, Komarneni S, Uchino K, "Influence of Composition and Dopant on Photostriction in Lanthanum-Modified Lead Zirconate Titanate Ceramics", J. Intelligent and Smart Mat. 10, (1999), p. 439–445.

7. Meng Q, Kui Y, Yung C L, "High efficient photovoltaics in nanoscaled ferroelectric thin films", Appl. Phys. Lett, 93, (2008), p. 122903-122904.

8. Meng Q, Kui Y, Yung C L, "Photovoltaic characteristics in polycrystalline and epitaxial $Pb_{0.97}La_{0.03}Zr_{0.52}Ti_{0.48}O_3$ ferroelectric thin films sandwiched between different top and bottom electrodes", J. Appl. Phys. 105, (2009), p. 061624-61631.

9. Qin M, Yao K, Liang Y C, Shannigrahi S, "Thickness effects on photoinduced current in ferroelectric $Pb_{0.97}La_{0.03}Zr_{0.52}Ti_{0.48}O_3$ thin films", J. Appl. Phys. 101, (2007), p. 014104-014108.

10. Nagaraj B, Aggarwal S, Song T K, Sawhney T, Ramesh R, "Leakage current mechanisms in lead-based thin-film ferroelectric capacitors", Physical Review B. 59, (1999), p. 16022–16027.

11. Okuyama M, Usuki T, Hamakawa Y, Nakagawa T, "Epitaxial growth of ferroelectric PLZT thin film and their optical properties", Applied Physics A: Materials Science & Processing. 21, (1980), p. 339-343.

12. Panda B, Ray S K, Dhar A, Sarkar A, Bhattacharya D, Chopra K L, "Electron beam deposited lead lanthanum zirconate titanate thin films for silicon based device applications", J. Appl. Phys. 79, (1996), p. 1008-1012.

13. Lakeman C D E, Payne D A, "Sol-gel processing of electrical and magnetic ceramics, Materials Chemistry and Physics" 38, (1994), p. 305-324.

14. Sayer M, Yi G, Sedlar M, "Comparative Sol gel processing of PZT thin films", Integrated Ferroelectrics. 7, (1995), p. 247-258.

15. Lange F F, "Chemical solution routes to single-crystal thin films", Science. 273, (1996), p. 903-909.

16. Zheng F, Xu J, Fang L, Shen M, Wu X, "Separation of the Schottky barrier and polarization effects on the photocurrent of Pt sandwiched Pb $Zr_{0.20}Ti_{0.80}O_3$ films", Appl. Phys. Lett. 93, (2008), p. 172101-172103.

17. Yang S Y, Seidel J, Byrnes S J, Shafer P, Yang C H, Rossell M D, Yu P, Chu Y H, Scott J F, Ager J W, Martin L W, Ramesh R, "Above-bandgap voltages from ferroelectric photovoltaic devices", Nature Nanotechnology. 5, (2010), p. 143-147.

18. D. Dimos, W. L. Warren, M. B. Sinclair, B. A. Tuttle, and R. W. Schwartz, "Photo induced hysteresis changes and optical storage in $(Pb,La)(Zr,Ti)O_3$ thin films and ceramics", J. Appl. Phys. 76, (1994), p. 4305–4315.

19. Kholkin A L, Iakovlev S O, Baptista J L, "Direct effect of illumination on ferroelectric properties of lead zirconate titanate thin films", Appl. Phys. Lett. 79, (2001), p. 2055-2057.

20. Dimos D, Potter B G, Sinclair M B, Tuttle B A, Warren W L, "Photo-induced and electrooptic properties of $(Pb,La)(Zr,Ti)O_3$ films for optical memories",Integrated Ferroelectrics. 5, (1994), p. 47-58.

21. Kholkin A L, Setter N, "Photoinduced poling of lead titanate zirconate thin films", Appl. Phys. Lett. 71, (1997), p. 2854-2856.

22. Harshan V N, Kotru S, "Effect of Annealing on Ferroelectric Properties of Lanthanum Modified Lead Zirconate Titanate Thin Films", Integrated Ferroelectrics. 130, (2011), p. 73-83.

23. Han H, Song X Y, Zhong J, Kotru S, Padmini P, Pandey R K, "Highly a-axis-oriented Nb-doped $Pb(Ti_xZr_{1-x})O_3$ thin films grown by sol-gel technique for uncooled infrared detectors", Appl. Phys. Lett. 85, (2004), p. 5310-5312.

24. Yang S Y, Martin L W, Byrnes S J, Conry T E, Basu S R, Paran D, Reichertz L, Ihlefeld J, Adamo C, Melville A, Chu Y H, Yang C H, Musfeldt J L, Schlom D G,

Ager J W, Iii, Ramesh R, "Photovoltaic effects in $BiFeO_3$", Appl. Phys. Lett. 95, (2009), p. 062903-062909.

25. Harshan V N, Kotru S, "Influence of work-function of top electrodes on the photovoltaic characteristics of $Pb_{0.95}La_{0.05}Zr_{0.54}Ti_{0.46}O_3$ thin film capacitors", Appl. Phys. Lett. 100, (2012) , p. 173901-173904.

26. Brody P. S, "Dielectric constant decrease upon Illumination in a Barium Titanate Crystal", Army Research Laboratory, Report , ARL-TR-905, (1997),  p.19970501.

27. Warren W L, Dimos D, Tuttle B A, E.Pike G, Raymond M V, D.Nashby R, Ramesh R, Evans J T, "Mechanism(s) for the supression of switchable polarization in PZT and $BaTiO_3$ ",MRS, 361, Cambridge University Press,(1994), p. 51.

28. Qin M, Yao K, Liang Y C, Gan B K, "Stability of photovoltage and trap of light-induced charges in ferroelectric $WO_3$-doped $Pb_{0.97}La_{0.03}Zr_{0.52}Ti_{0.48}O_3$ thin films", Appl. Phys. Lett. 91, (2007), p. 92904–092903.

# CHAPTER 4

# INFLUENCE OF WORK-FUNCTION OF TOP ELECTRODES ON THE PHOTOVOLTAIC CHARACTERISTICS OF $Pb_{0.95}La_{0.05}Zr_{0.54}Ti_{0.46}O_3$ THIN FILM CAPACITORS [3]

**Abstract**

Photovoltaic properties of $Pb_{0.95}La_{0.05}Zr_{0.54}Ti_{0.46}O_3$ thin film capacitors prepared using solution based method with metal top electrodes having different work functions are investigated in this work. It is shown that by using aluminum, a low work-function metal, as top electrode, the magnitude of photo voltage as well as photo current can be enhanced. More than one magnitude enhancement in the photovoltaic efficiency is observed with Al as top electrode compared to Pt electrodes. This work clearly highlights that an appropriate choice of low work function metal electrode can enhance the photovoltaic response of the ferroelectric thin film capacitors.

---

Ferroelectric materials have attracted considerable attention in recent years as a potential photovoltaic (PV) material both in bulk and as well as in thin film form [1-4]. These materials do not yet exhibit high conversion efficiency as compared to the conventional semiconductor based solar devices. Recent observation of achieving above band-gap open circuit voltages from such materials, along with the ability to tune the photovoltaic response by controlling the ferroelectric polarization, have generated major interest towards realizing non-conventional solar devices [3,5,6]. It is widely agreed that the mechanism for the PV effect in a ferroelectric material is based on photo-generated electron-hole pairs which is separated by the internal electric field arising from the polarization of the ferroelectric [4,7], as observed in a variety of ferroelectric oxides viz. bismuth ferrite ($BiFeO_3$), lithium niobate ($LiNbO_3$), Barium titanate ($BaTiO_3$) and La based lead zirconate titanate (PLZT) [2,4,8,9]. PLZT is particularly interesting material for devices because of its optical transparency and lower processing temperature. The compatibility of PLZT films with plastics, transparent substrates as well as with transparent conductors gives an added advantage to harness energy at the lower visible spectrum in addition to UV sensor applications [10,11].

Control of ferroelectric polarization has been reported as a way to increase the PV efficiency of devices based on ferroelectric thin films of $BiFeO_3$ [2,3] and PLZT [4-6] . Apart from polarization tuning, modification of ferroelectric-electrode interface is a promising way to increase the efficiency of the PV devices [9,12,13]. Recently it was observed that the photo-voltaic efficiency is substantially influenced by the interface field arising from the ferroelectric-electrode Schottky contact [12,13]. This interface-field is polarization-independent unlike the bulk depolarization field. Therefore, one of the ways to change the photo-conductivity of the ferroelectric based thin film devices is choosing appropriate electrodes to affect the space charge densities at the ferroelectric-metal interface [13,14].

In general, ferroelectrics exhibit the J-V characteristics of a back-to-back Schottky diode in a capacitor configuration. The Schottky barrier at the metal-ferroelectric interface exists due to the difference between the work functions, which creates a polarization-independent interface electric field. Effects of such interfacial field on the photo-conductivity of the PZT films have been reported [14-16]. The work by Qin et al.[13] emphasizes the importance of the dielectric constants of the electrode material, whereas other reports point to the importance of asymmetrical Schottky contacts [9,12,14,17].

The purpose of this work is to investigate the influence of work-function of the top electrodes on photovoltaic properties of La doped PZT based thin film capacitors. PV results obtained from two types of capacitor devices with metal electrodes (Pt and Al) having different work functions are compared. For this work, PLZT sol (composition $Pb_{0.95}La_{0.05}Zr_{0.54}Ti_{0.46}O_3$ was prepared using solution-based technique [18] and the films were deposited on commercially obtained Pt/Si wafers by spin coating method. The choice of La doping and the Zr/Ti ratio was motivated by the work of Poosanas et al [19] , where this particular composition (5/54/46) was reported to produce the highest photo-voltage. The films were approximately 210 nm thick (measured using profilometry) and were post-annealed in flowing oxygen using a RTA (Qualiflow) at 750$^o$C. Details on the effect of processing temperatures on ferroelectric and structural properties are discussed elsewhere [20]. The films were characterized using X-ray diffraction (Rigaku) and surface roughness was estimated using a commercial Atomic Force Microscope (Digital Instruments 3100).

These films were further used to prepare metal/PLZT/metal capacitor structures by vacuum deposition of top electrode using a shadow mask. Circles with an area of $9.1 \times 10^{-8} m^2$ and thickness of 70 nm were deposited to form the top electrodes (Platinum and Aluminum) and Platinum substrate served as bottom electrode. Thus in both cases, the bottom interface remains the same whereas the top metal-ferroelectric interface is influenced by changing the top electrodes. Capacitance of the stack was measured using a probe station (Signatone) connected to a LCR meter (4284A from HP). Ferroelectric measurements on the films were obtained using a commercial precision workstation from Radiant Technologies. All measurements were taken using the top-bottom capacitor configuration. The existing probe station was modified into a photovoltaic testing unit with provisions to illuminate the sample from the top. A visible lamp source (SCI-200 Science Tech) was used in this work. The intensity of the light source was measured and calibrated using a meter (Newport Corporation, 1918-C). A Keithley (6517A) meter was used to measure J–V characteristics, which was interfaced with a computer for data acquisition using LabView® software.

Figure 4.1 shows the structural characterization of the film grown on platinized silicon substrate. The film exhibits polycrystalline behavior, similar to previously reported films grown by a similar technique [6] . Roughness of the films annealed at 750°C was measured to be 7.6 nm (inset of figure 1). Dark and illuminated J-V curves of Pt/PLZT/Pt capacitor are shown in figure 4(a). These curves are linear for different polarization states (positive, unpoled and negative), with a trend consistent with the report of Qin et al [6] for PLZT (3/52/48) films with Au top electrodes. Open-circuit voltage ($V_{oc}$) for the unpoled sample was 0.17 V and it changed to 0.12 V after applying a positive poling of 5 volts and to 0.24 V after applying a negative poling of same magnitude. The short circuit current density ($J_{sc}$) of the device was -6.44 x $10^{-7}$ A/cm$^2$, -4.06 x$10^{-7}$ A/cm$^2$, and -7.88 x $10^{-7}$ A/cm$^2$ for unpoled, positively poled, and negatively poled samples, respectively. From the J-V data of figure 2(a), the power conversion efficiency of the device was calculated using the integral input power density (48 mW/cm$^2$), measured over the spectral range of 250 nm to 1100 nm.

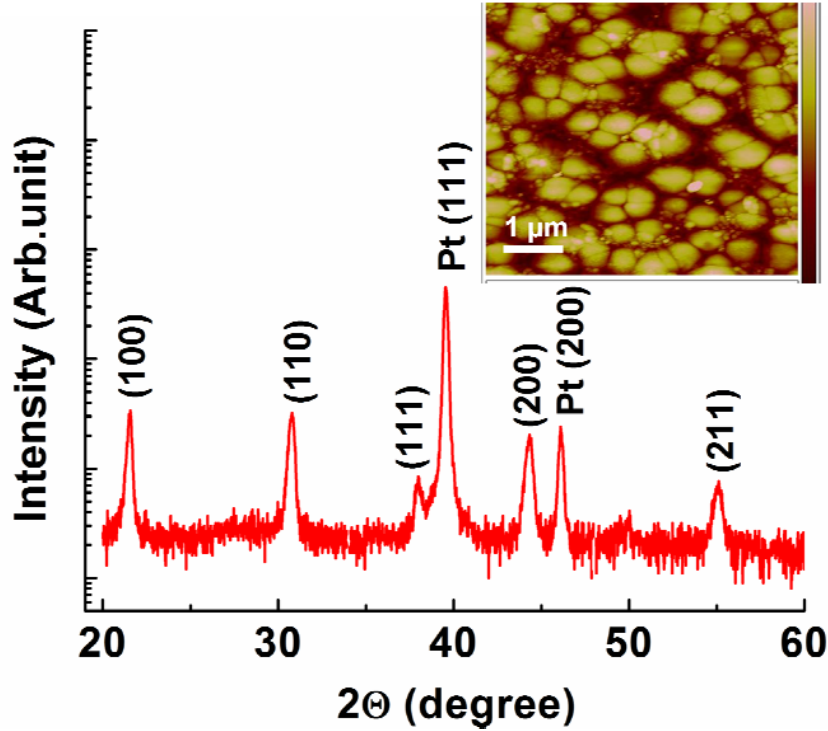

**Figure 4.1**: X-ray diffraction of sol gel grown PLZT thin film showing polycrystalline texture, Inset shows the AFM micrograph of the film.

From figure 4.2 (b), the maximum power conversion efficiency in this spectral range is found to be 1.03x $10^{-4}$ % for the negatively poled films. Owing to the large band gap of PLZT, photons with energy smaller than the band gap such as the visible light used in this study, do not fully contribute to the PV response of the films. Usage of monochromatic light source corresponding to the band gap of the films is expected to further improve the efficiency of the films. The shift of J-V curves with poling is similar to the report of Qin et al [6] where the shift is reported to be due to the depolarization field of the bulk ferroelectric. Existence of such a depolarization field affecting the photo-conductivity of sol-gel based PZT thin films are also reported by Kholkin et al [21] and Fengang et al [22] .

Figure 4.3, shows PE curves of the PLZT film capacitors with two different top electrodes (Pt and Al) and identical Pt bottom electrodes. Both configurations showed the expected ferroelectric switching loop behavior indicating ferroelectric properties. With Pt as the top electrode, the film shows a symmetrical loop with saturation polarization ($P_s$) of 72 µC/cm$^2$, a remnant polarization ($P_r$) of 23 µC/cm$^2$ and an approximate coercive field ($E_c$) of 25 kV/cm. With Al as the top electrode, an asymmetrical loop was observed having lower values of saturation polarization ($P_s$) and remnant polarization ($P_r$) (44 µC/cm$^2$ and 18 µC/cm$^2$, respectively), and a higher coercive field ($E_c$) (175 kV/cm). Such a change in the shape of the hysteresis loop can be attributed to a possible formation of a thin $Al_2O_3$ layer (a few nm) at the interface, owing to the tendency of Al to oxidize. Similar effects on hysteresis shapes were observed by Ma et al [23] on PZT/$Al_2O_3$ bilayers.

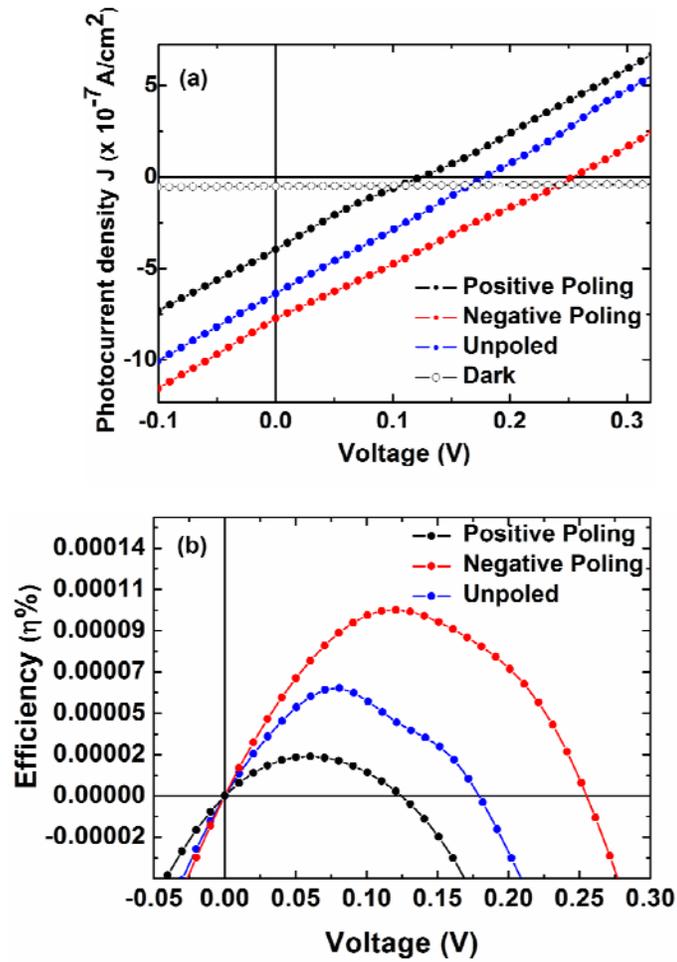

**Figure 4.2** (a): J-V behavior of Pt/PLZT/Pt capacitors under different poling conditions (b) Power conversion efficiency calculated from J-V curves using spectral intensity of 48mW/cm$^2$

Figures 4.4 (a & c) show the illuminated J-V characteristics and efficiency of the PLZT capacitor type devices with two distinct top electrodes. A schematic of the energy band diagram of the device is shown in figure 4(b). The PLZT film has a band gap of 3.8 eV as measured using spectroscopic technique (results not included here) and is considered to be an n type material [24]. Work functions of Al and Pt are 4.2 eV and 5.3 eV respectively [25,6]. Since the Pt bottom electrode has a higher work function than PLZT, a Schottky barrier is formed at the Pt/PLZT interface with an electric field $E_{in}$. With illumination, $E_{in}$ and the bulk depolarization field $E_{dp}$ together contribute to the electric output of the device. It is observed that when Pt is used as a top electrode, the $V_{oc}$ and the $J_{sc}$ for the device are 0.17 V and $-6.44 \times 10^{-7}$ A/cm$^2$ respectively. Both $V_{oc}$ and the $J_{sc}$ are seen to increase to 0.37 V and $-3.6 \times 10^{-6}$ A/cm$^2$ with Al as the top electrode.

The power conversion efficiency of the device is seen to increase by an order of magnitude, from $6.31 \times 10^{-5}$ % to $7.08 \times 10^{-4}$ % by replacing Pt with Al as a top electrode. The order of magnitude increase in the power conversion efficiency is attributed to the lower work function of the Al top electrodes. It is pertinent to point out here that since the work function of $Al_2O_3$ is lower or comparable to pure Al [25,26], a potential presence of a very thin layer of $Al_2O_3$ between Al and PLZT, might have further contributed to the enhancement of the photo-current.

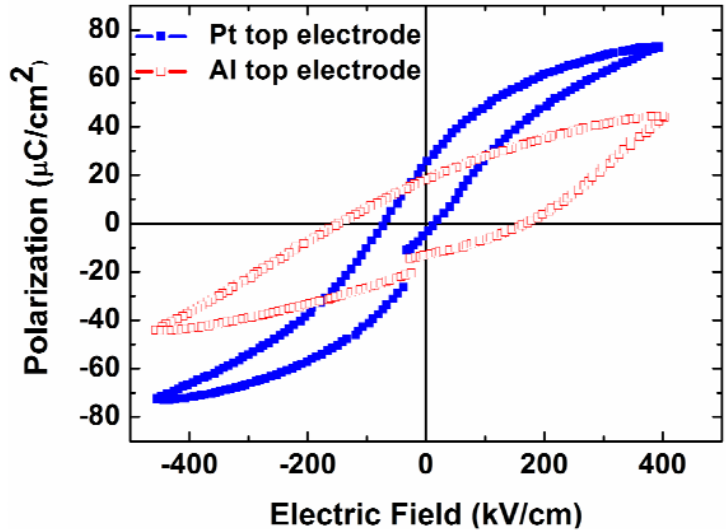

**Figure 4.3**: Polarization vs Electric field (P-E) behavior of PLZT films with Pt and Al top electrodes

It has been shown that distribution of screening charges at the electrodes influences the photo conductivity of the ferroelectric films. Further, the use of oxide electrodes, with higher dielectric constants, has been reported to increase the efficiency of the ferroelectric PV devices [13]. Recent reports by Pintilie [15,27] and Gan [11] also point to a profound influence on the PV characteristics of ferroelectric thin films with change in the work functions of the top electrodes. The observed increase in the open circuit voltage and the short circuit current of the devices is attributed to the asymmetric interface field caused by electrodes of different work functions. Influence of such an asymmetric interface field has also been recently shown by Gan et al [11], where a conducting oxide (LSMO) was used as the top electrode.

There are various other factors which are reported to induce asymmetry in the ferroelectric metal interface, such as non-uniform distribution of charge effects, mechanical stress by the substrate, thermal annealing, charge injection, and accumulation during polarization switching [28-31]. In our case, to minimize the measurement variation, both Pt and Al top electrodes were deposited on the same PLZT film. Thus we believe that the influence of these factors contributing to the asymmetry should be similar for Pt and Al top electrodes, and can be considered a secondary effect. Hence the increased PV response from the films is considered to arise from the difference in the work functions of the top electrodes.

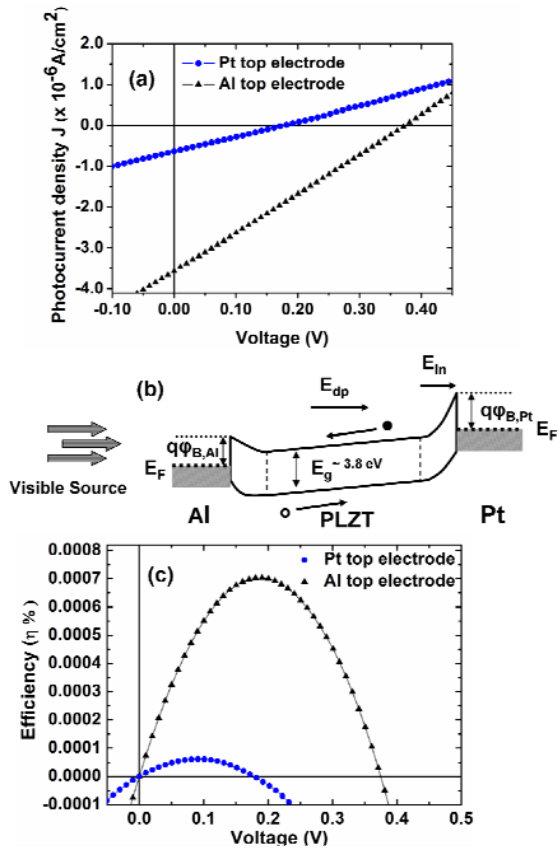

**Figure 4.4** (a): Illuminated J-V behavior of unpoled PLZT films with Pt and Al top electrodes.(b) The energy band diagram of the Al/PLZT/Pt capacitor configuration.(c) Power conversion efficiency calculated from J-V curves of fig 4 (a) using spectral intensity of 48mW/cm$^2$.

To summarize, PLZT (Pb$_{0.95}$La$_{0.05}$Zr$_{0.54}$Ti$_{0.46}$O$_3$) films were prepared using a solution-based technique. These films were used to fabricate metal/PLZT/metal capacitor structures. The photovoltaic response was investigated using two distinct top electrodes. With Al as a top electrode, the illuminated J-V curves of the device significantly improved and the open circuit voltage increased from 0.17 V to 0.37 V. This resulted in an enhancement of PV efficiency by one order of magnitude. The improved efficiency of Al/PLZT/Pt devices is attributed to the lower work function of the top Al electrode.

## 4.1   Acknowledgments

The work was supported by the National Science Foundation under ECCS Grant No. 0943711 and by the startup funds provided by the College of Engineering, UA. The authors extend their thanks to the Center for Materials and Information Technology (MINT), University of Alabama, for providing the characterization facilities. Prof. Gregory J. Szulczewski, Department of Chemistry, UA is acknowledged for help with e-beam evaporator.

## 4.2 Chapter 4 References

1    T. Choi, S. Lee, Y. J. Choi, V. Kiryukhin, and S. W. Cheong, "Switchable Ferroelectric Diode and Photovoltaic Effect in BiFeO3," Science, 324 (5923), (2009), p. 63-66.

2    S. Y. Yang, L. W. Martin, S. J. Byrnes, T. E. Conry, S. R. Basu, D. Paran, L. Reichertz, J. Ihlefeld, C. Adamo, A. Melville, Y. H. Chu, C. H. Yang, J. L. Musfeldt, D. G. Schlom, J. W. Ager, Iii, and R. Ramesh, "Photovoltaic effects in $BiFeO_3$," Applied Physics Letters, 95 (6), (2009), p. 062909.

3    S. Y. Yang, J. Seidel, S. J. Byrnes, P. Shafer, C. H. Yang, M. D. Rossell, P. Yu, Y. H. Chu, J. F. Scott, J. W. Ager, L. W. Martin, and R. Ramesh, "Above-bandgap voltages from ferroelectric photovoltaic devices," Nature Nanotechnology, 5 (2), (2010), p. 143-147.

4    Q. Meng, Y. Kui, and C. L. Yung, "High efficient photovoltaics in nanoscaled ferroelectric thin films," Applied Physics Letters, 93 (12), (2008), p. 122904.

5    M. Qin, K. Yao, Y. C. Liang, and B. K. Gan, "Stability of photovoltage and trap of light-induced charges in ferroelectric $WO_3$-doped $Pb_{0.97}La_{0.03}Zr_{0.52}Ti_{0.48}O_3$ thin films," Applied Physics Letters, 91 (9), (2007), p. 092904.

6    Q. Meng, Y. Kui, and C. L. Yung, "Photovoltaic characteristics in polycrystalline and epitaxial $Pb_{0.97}La_{0.03}Zr_{0.52}Ti_{0.48}O_3$ ferroelectric thin films sandwiched between different top and bottom electrodes," Journal of Applied Physics, 105 (6), (2009), p. 061624.

7    H. T. Huang, "Solar energy: Ferroelectric photovoltaics," Nature Photonics, 4 (3), (2010), p. 134-135.

8    A. M. Glass, D. V. D. Linde, and T. J. Negran, "High voltage bulk photovoltaic effect and the photorefractive process in $LiNbO_3$," Applied Physics Letters, 25 (4), (1974), p. 233-235.

9	Y. S. Yang, S. J. Lee, S. Yi, B. G. Chae, S. H. Lee, H. J. Joo, and M. S. Jang, "Schottky barrier effects in the photocurrent of sol--gel derived lead zirconate titanate thin film capacitors," Applied Physics Letters, 76 (6), (2000), p. 774-776.

10	G. Bee Keen, Y. Kui, L. Szu Cheng, C. Yi Fan, and G. Phoi Chin, "An Ultraviolet (UV) Detector Using a Ferroelectric Thin Film With In-Plane Polarization," Electron Device Letters, IEEE, 29 (11), (2008), p. 1215-1217.

11	G. Bee Keen, Y. Kui, L. Szu Cheng, G. Phoi Chin, and C. Yi Fan, "A Ferroelectric Ultraviolet Detector With Constructive Photovoltaic Outputs," Electron Device Letters, IEEE, 32 (5), (2011), p. 665-667.

12	V. Yarmarkin, B. M. Goltsman, M. Kazanin, and V. Lemanov, "Barrier photovoltaic effects in PZT ferroelectric thin films," Physics of the Solid State, 42 (3), (2000), p. 522-527.

13	M. Qin, K. Yao, and Y. C. Liang, "Photovoltaic mechanisms in ferroelectric thin films with the effects of the electrodes and interfaces," Applied Physics Letters, 95 (2), (2009), p. 022912-022913.

14	J. Xu, D. Cao, L. Fang, F. Zheng, M. Shen, and X. Wu, "Space charge effect on the photocurrent of Pt-sandwiched $Pb(Zr_{0.20}Ti_{0.80})O_3$ film capacitors," Journal of Applied Physics, 106 (11), (2009), p. 113705-113705.

15	L. Pintilie, C. Dragoi, and I. Pintilie, "Interface controlled photovoltaic effect in epitaxial $Pb(Zr,Ti)O_3$ films with tetragonal structure," Journal of Applied Physics, 110 (4), (2011), p. 044105-044106.

16	L. Pintilie, I. Vrejoiu, G. Le Rhun, and M. Alexe, "Short-circuit photocurrent in epitaxial lead zirconate-titanate thin films," Journal of Applied Physics, 101 (6), (2007), p. 064109-064108.

17	D. Cao, J. Xu, L. Fang, W. Dong, F. Zheng, and M. Shen, "Interface effect on the photocurrent: A comparative study on Pt sandwiched $Bi_{3.7}Nd_{0.3}Ti_3O_{12}$ and $PbZr_{0.2}Ti_{0.8}O_3$ films," Applied Physics Letters, 96 (19), (2010), p. 192101-192103.

18  H. Han, X. Y. Song, J. Zhong, S. Kotru, P. Padmini, and R. K. Pandey, "Highly a-axis-oriented Nb-doped Pb(Ti$_x$Zr$_{1-x}$)O$_3$ thin films grown by sol-gel technique for uncooled infrared dectors," Applied Physics Letters, 85 (22), (2004), p. 5310-5312.

19  P. Poosanaas, K. Tonooka, I. R. Abothu, S. Komarneni, and K. Uchino, "Influence of Composition and Dopant on Photostriction in Lanthanum-Modified Lead Zirconate Titanate Ceramics," Journal of Intelligent Material Systems and Structures, 10 (6), (1999), p. 439-445.

20  V. N. Harshan and S. Kotru, "Effect of Annealing on Ferroelectric Properties of Lanthanum Modified Lead Zirconate Titanate Thin Films," Integrated Ferroelectrics, 130 (1), (2011), p. 73-83.

21  A. Kholkin, O. Boiarkine, and N. Setter, "Transient photocurrents in lead zirconate titanate thin films," Applied Physics Letters, 72 (1), (1998), p. 130-132.

22  F. Zheng, J. Xu, L. Fang, M. Shen, and X. Wu, "Separation of the Schottky barrier and polarization effects on the photocurrent of Pt sandwiched Pb Zr$_{0.20}$Ti$_{0.80}$O$_3$ films," Applied Physics Letters, 93 (17), (2008), p. 172101-172103.

23  Z. Ma and A. Q. Jiang, "Study of Ferroelectic Properties in Ferroelectric/High-k Dielectric Bilayers," Ferroelectrics, 401 (1), (2010), p. 129-133.

24  M. Qin, K. Yao, Y. C. Liang, and B. K. Gan, "Stability and magnitude of photovoltage in ferroelectric Pb$_{0.97}$La$_{0.03}$Zr$_{0.52}$Ti$_{0.48}$O$_3$ thin films in multi-cycle UV light illumination," Integrated Ferroelectrics, 95, (2007), p. 105-116.

25  Y.-C. Yeo, T.-J. King, and C. Hu, "Metal-dielectric band alignment and its implications for metal gate complementary metal-oxide-semiconductor technology," Journal of Applied Physics, 92 (12), (2002), p. 7266-7271.

26  V. V. Afanas'ev, M. Houssa, A. Stesmans, and M. M. Heyns, "Band alignments in metal--oxide--silicon structures with atomic-layer deposited Al$_2$O$_3$ and ZrO$_2$," Journal of Applied Physics, 91 (5), (2002), p. 3079-3084.

27  L. Pintilie, V. Stancu, E. Vasile, and I. Pintilie, "About the complex relation between short-circuit photocurrent, imprint and polarization in ferroelectric thin films," Journal of Applied Physics, 107 (11), (2010), p. 114111-114116.

28  P. J. Schorn, Bottger, Ulrich, Waser, Rainer, "Monte Carlo simulations of imprint behavior in ferroelectrics," Applied Physics Letters, 87 (24), (2005), p. 242902-242903.

29  A. K. Tagantsev, I. Stolichnov, N. Setter, and J. S. Cross, "Nature of nonlinear imprint in ferroelectric films and long-term prediction of polarization loss in ferroelectric memories," Journal of Applied Physics, 96 (11), (2004), p. 6616-6623.

30  A. Q. Jiang, Tang, T. A., "Congruent charge-injection spectrum from independent measurements of fatigue and imprint in ferroelectric thin films," Journal of Applied Physics, 105 (6), (2009), p. 061608-061604.

31  G. H. Kim, Lee, Hyun Ju, Jiang, an Quan, Park, Min Hyuk, Hwang, Cheol Seong, "An analysis of imprinted hysteresis loops for a ferroelectric Pb(Zr,Ti)$O_3$ thin film capacitor using the switching transient current measurements," Journal of Applied Physics, 105 (4), (2009), p. 044106-044105.

# CHAPTER 5

# IRRADIANCE DEPENDENT EQUIVALENT MODEL FOR PLZT BASED FERROELECTRIC PHOTO VOLTAIC DEVICES [4]

**Abstract**

Ferroelectric materials have been known for decades but the photovoltaic response of ferroelectric materials is a relatively new area of research and has become a fast emerging field owing to the fact that high values of open circuit voltages are possible from solar cells made from such materials when compared to the conventional photovoltaic (PV) semiconductor device. Thus simulation study of predicting the I-V response with cell voltage and irradiance is of interest to the ferroelectric as well as electronic device community. In this work, the current–voltage (I-V) characteristics of ferroelectric photovoltaic cells based on $Pb_{0.95}La_{0.05}Zr_{0.54}Ti_{0.46}O_3$ (PLZT) films were studied and modeled as a function of irradiance. The current response at a given irradiance is a function of the solar cell voltage (V) and irradiance level (G), which is parameterized, modeled and found to fit approximately to a line equation. The experimentally measured I-V data were used to extract the parameters for the characteristic equation using a mathematical model approach, which was developed using MATLAB® software package. The derived equation was then used to simulate the I-V curves at unknown irradiance levels to predict the performance of the PLZT photovoltaic cell, and the simulated results were compared with the experimentally measured data. An

---

equivalent electric circuit model based on characteristic equation was developed to predict the behavior of PLZT PV device under various irradiance levels.

## 5.1 Introduction

Ferroelectric photovoltaic have attracted interest in the past few years due to the observation of high band-gap open circuit voltages from such devices. These materials are believed to be potential alternative to the conventional Photovoltaic (PV) semiconductor technology [1]-[3]. Such devices typically have capacitor configuration with an active ferroelectric layer embedded between two electrodes, similar to metal–insulator–metal type devices. Among the many available ferroelectrics, few materials such as bismuth ferrite ($BiFeO_3$) and La based Lead Zirconate Titanate (PLZT) have been reported to show the photovoltaic capabilities [1],[2]. PLZT based films have additional advantage for such devices due to their optical transparency which makes them particularly attractive for applications towards sensors devices and photo-diodes at the ultra violet as well as visible region [4],[5]. Recent work on photovoltaic properties of PLZT devices show that short circuit current densities ($J_{sc}$) for such devices are in range of $nA/cm^2$ and it is possible to achieve the power conversion efficiencies as high as 0.3% with oxide top electrodes [6]. The ability to control the short circuit current density ($J_{sc}$) and open circuit voltage ($V_{oc}$) with externally applied electrical poling in ferroelectric devices makes these materials promising for applications towards non-conventional solar devices [7].

Photovoltaic research in a broader sense can be divided into distinct areas such as device physics and fabrication, power conversion system design and maximum power point tracking control techniques [8]-[17]. This work is concerned with development of a mathematical and circuit model that is useful for simulating and predicting the device performance of a PLZT based solar cell and in power electronics applications. The area of ferroelectric photovoltaic is relatively new and behavior of devices made from such materials is yet to be fully understood. There are a few reports on the photovoltaic behavior of devices with varying light irradiance [18],[19]. Understanding these devices in terms of I-V parameters based on varying irradiance is essential to further understand and interpret the PV mechanism in such devices. A mathematical model, based on experimentally measured data, was used to simulate and predict the performance of the ferroelectric photovoltaic devices. Among the various modeling approaches, modeling PV cell with equivalent circuit [20], and curve fitting approach is widely used to develop mathematical models [21],[22]. The curve fitting approach has been used first in the present work and then an electric circuit model is developed. The simulation and modeling studies on irradiance dependent electrical characteristics of $Pb_{0.95}La_{0.05}Zr_{0.54}Ti_{0.46}O_3$ (PLZT) based ferroelectric PV cells are reported in this work. Current–Voltage (I-V) characteristics were measured at various irradiance levels. The input parameters were chosen to be irradiance level (G) and cell operating voltage (V). These two parameters were used to identify the output characteristic I-V equation for that particular cell modeled by Simulink® tool. The characteristic equation thus derived, was later used to predict the I-V behavior of the cells at different irradiance levels.

## 5.2  Experimental Measurements

The device was fabricated by sandwiching a thin film of $Pb_{0.95}La_{0.05}Zr_{0.54}Ti_{0.46}O_3$ (PLZT) between two platinum electrodes. PLZT was fabricated using sol-gel method on Pt wafers and the electrodes were deposited by sputtering. For this work films with 210 nm thickness annealed at 550°C were chosen to fabricate devices. Further details about the fabrication process of the PLZT films and fabrication of device structure are reported elsewhere [23]. All the I-V measurements were taken with a top-bottom capacitor configuration with area of the fabricated capacitor being $9.1 \times 10^{-8}\,m^2$. A simple schematic of the photovoltaic test setup used for these measurements along with the cartoon of the device fabricated is shown in Fig. 5.1. The setup has the capability to illuminate the device from top. A visible lamp source was used for this work and the irradiance of the light was measured and calibrated using a meter obtained commercially (Newport 1918-C). A Keithley (6517) source meter was used to measure I-V curves, by sweeping voltage to measure current. The setup was interfaced with a computer for data acquisition using LabView®.

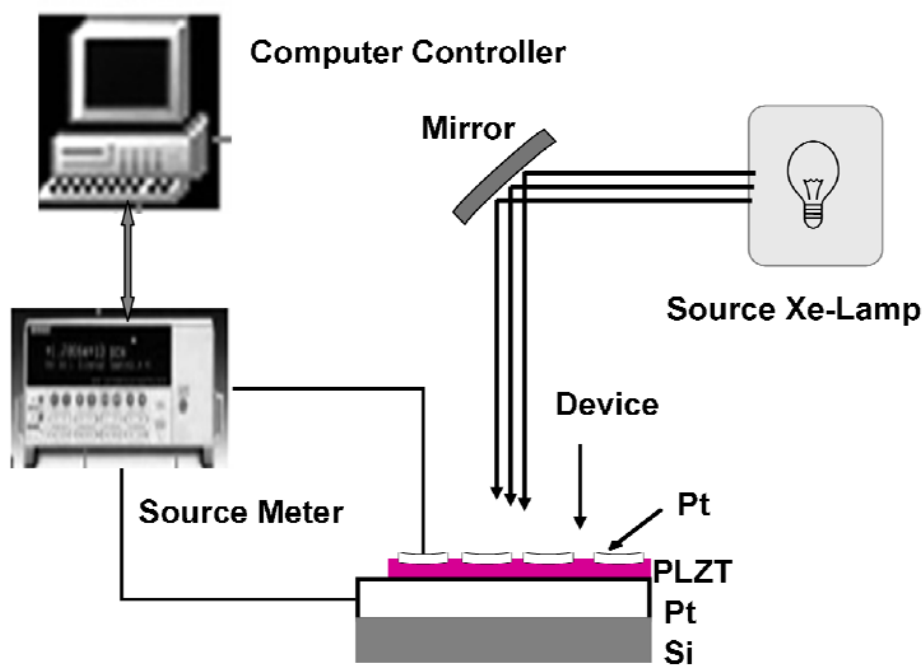

**Fig. 5.1** Schematic of the measurement setup used to measure the (I-V) data.

**5.3** **Results: Modeling and Discussion**

Current–Voltage (I-V) characteristics were measured at various irradiance levels from 110-140 mW/cm$^2$. Fig. 5.2 shows the experimentally measured data, showing a straight line trend, similar to I-V behavior observed for other ferroelectric PV devices [1],[18],[19]. Such a trend is believed to arise from the mechanism of charge carrier transport, which is different for ferroelectrics when compared to the semiconductor based PV devices [3].

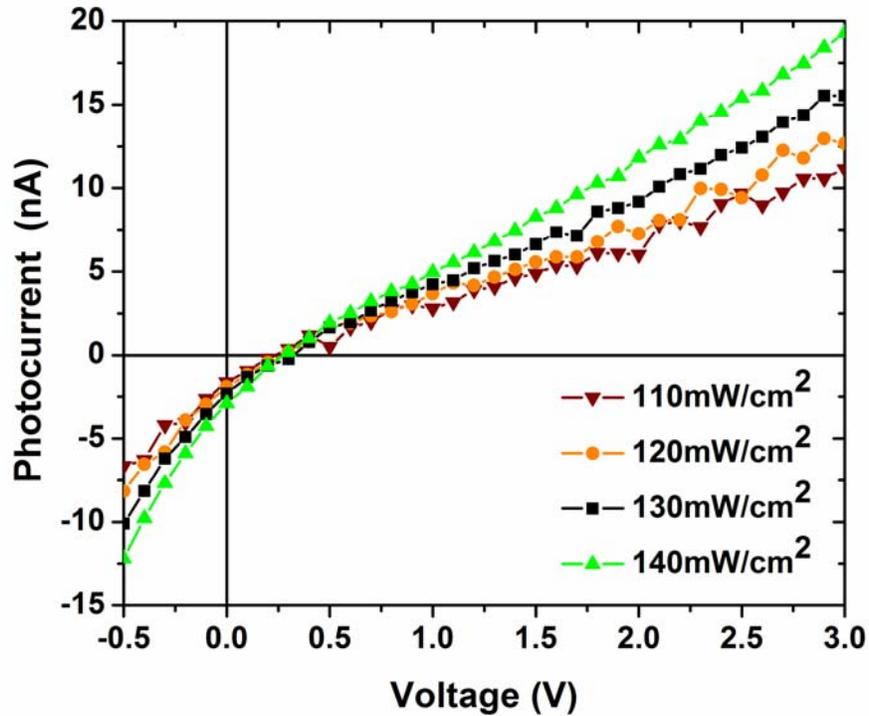

**Fig. 5.2** Experimentally measured I-V curves of PLZT photovoltaic cell under various irradiance levels

It is observed that at higher irradiance levels, the noise level in the data is reduced, which is expected as more charge carriers are generated at higher irradiance levels. The data measured under irradiance of 120 mW/cm² and 130 mW/cm² was used to extract the input parameters for the mathematical model, whereas data measured under irradiance of 110 mW/cm² and 140 mW/cm² was used to validate the model.

For the mathematical model, irradiance (G) and the input voltage (V) were chosen as the input parameters. Based on the experimental I-V data of the PV cell (Fig. 5.2) for the irradiance of 130mW/cm² and 120mW/cm², the current and voltage (I-V) behavior of a single PLZT PV cell was extracted and the fit is shown in Fig. 5.3 (a and b). From Fig. 5.3 (a), and the fit, the equation for the current (I) for the PV cell under the 130mW/cm² irradiance as a function of the voltage (V) was found to be

$$I = (5.7 \times 10^{-9}) \cdot V - 1.7 \times 10^{-9} \tag{1}$$

Similarly, equation for the current derived from fit as seen from Fig 5.3 (b), under irradiance of 120mW/cm² as function of voltage (V) was found to be

$$I = (4.6 \times 10^{-9}) \cdot V - 1.3 \times 10^{-9} \tag{2}$$

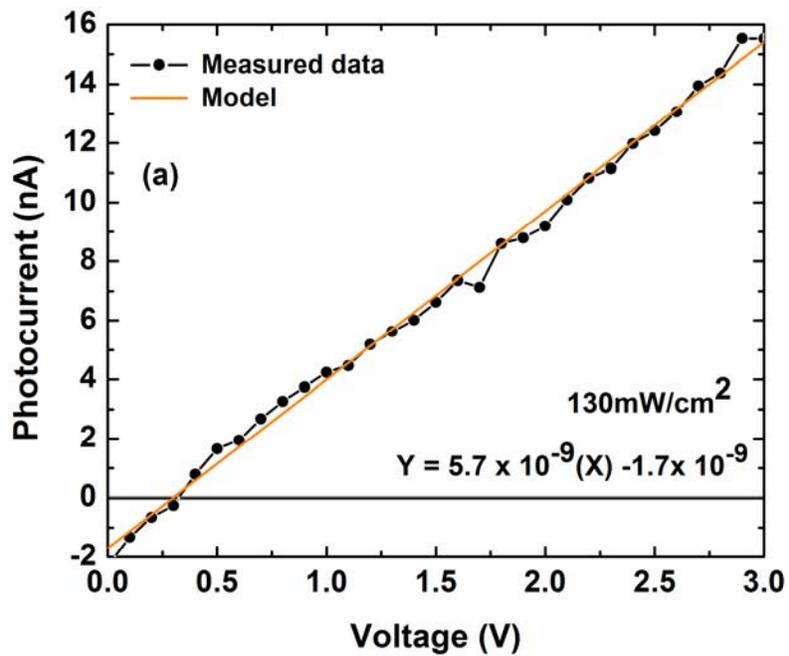

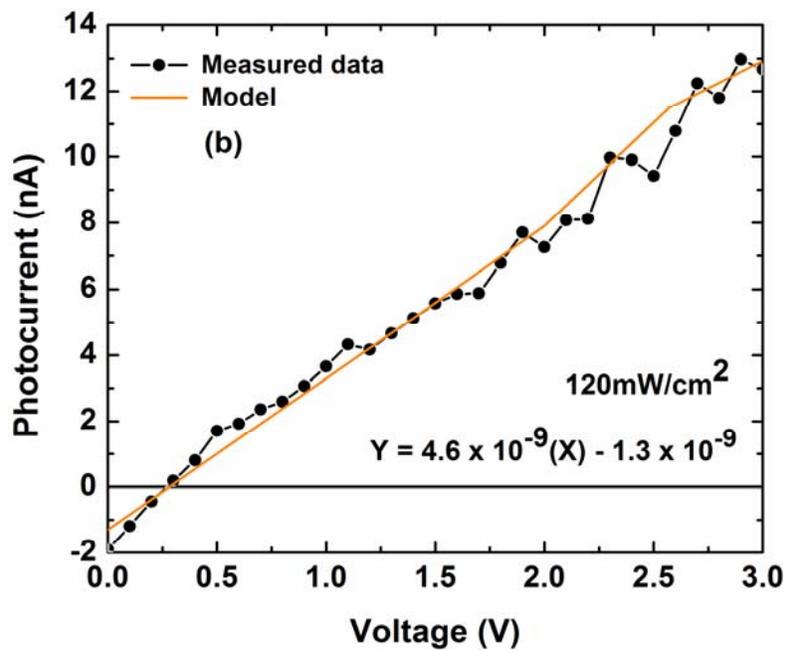

**Figs. 5.3** Simulated results of I-V curves obtained from experimental data for irradiance of (a) 130mW/cm$^2$ and (b) 120mW/cm$^2$.

For equations (1 and 2), V is the input voltage for the cell. The open circuit voltage ($V_{oc}$) can be calculated by setting the current (I) to zero. It is observed that the $V_{oc}$ changes with the input irradiance, which is also one of the characteristics of ferroelectric PV devices [14]. Using the above equations (1 and 2), a complete model based as a function of irradiance level (G) and the cell voltage (V) is obtained and is represented by equation (3).

$$\begin{aligned} I &= (1.62 \times 10^{-10} \cdot G - 9.8 \times 10^{-9}) \cdot V \\ &\quad + (-4 \times 10^{-11} \cdot G + 3.5 \times 10^{-9}) \end{aligned} \quad (3)$$

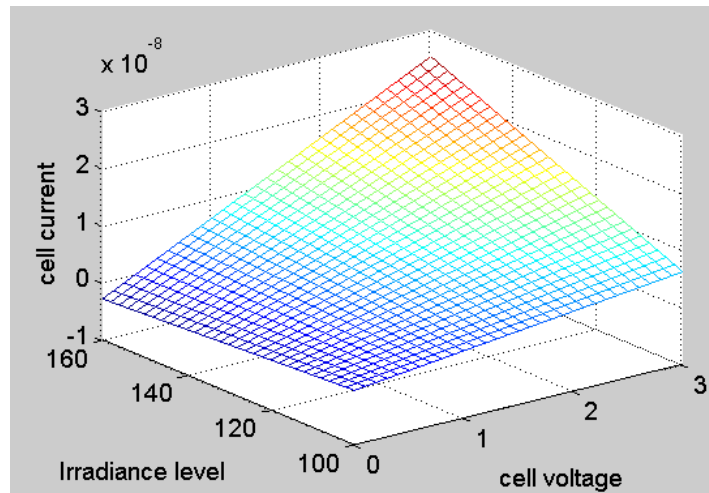

**Fig. 5.4.** The 3D plot of the photocurrent predicted in the range for irradiance (G) from (110-160 mW/cm$^2$) and cell voltage from (0-3 V) using Simulink®/MATLAB® software package.

Equation (3) is the behavioral current equation of the model with the inputs being assigned to irradiance level (G) and the cell voltage (V). Based on the equation (3), a 3-D plot of the current for the range of irradiance (G) from (110-140 mW/cm$^2$ ) and cell voltage from (0-3V) is shown in Fig. 5.4 and a snap shot of the simulation model used for PV device, realized using Simulink®/MATLAB® software package is shown in Fig. 5.5 and Fig. 5.6.

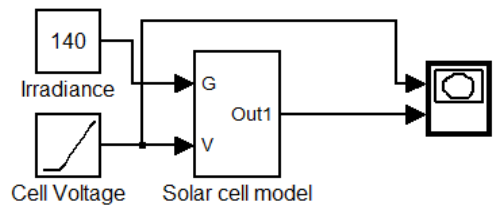

(a)

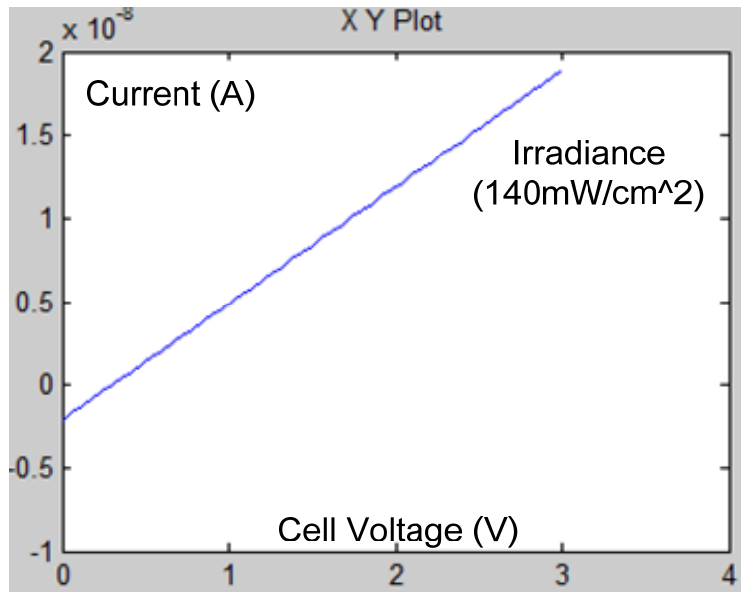

(b)

**Figs. 5.5.** (a) Snapshot of model realization using Simulink®/MATLAB® tool in MATLAB® based on equation (3) (b) cell current with cell voltage sweep under 140mW/cm$^2$ irradiance.

Fig. 5.5 (a) and Fig. 5.6 (a) show the cell model with two inputs (cell voltage and irradiance level). Fig. 5.5 (b), shows with simulated cell current with cell voltage swept from 0V to 3V keeping the irradiance level constant, at 140mW/cm². Fig. 5.6 (b), shows the simulated cell current with irradiance level swept from 100mW/cm² to 150mW/cm² with constant cell voltage of 1V. The behavioral equation (3) and the cell model from Fig. 5.5(a) and Fig. 5.6(a) were used to predict the photocurrent (I) for irradiance of 140 mW/cm² and 110 mW/cm², which is shown in Fig. 5.7 (a and b).

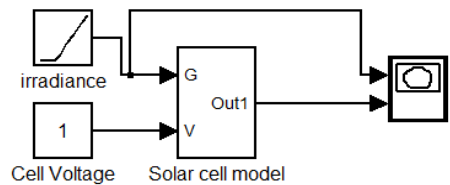

(a)

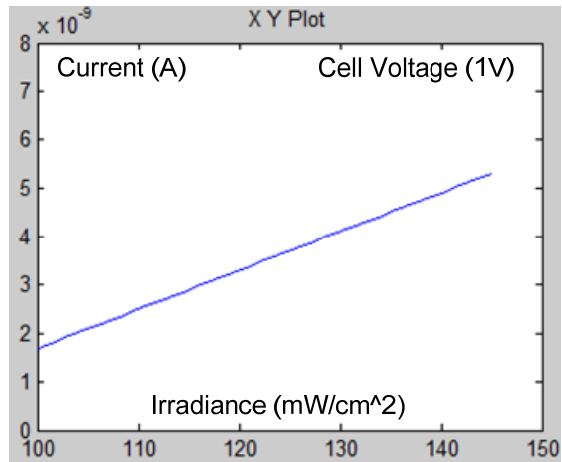

(b)

**Figs. 5.6.** (a) Snapshot of model realization using Simulink®/MATLAB® tool in MATLAB® based on equation (3) (b) cell current with irradiance level sweep under 1V operation voltage.

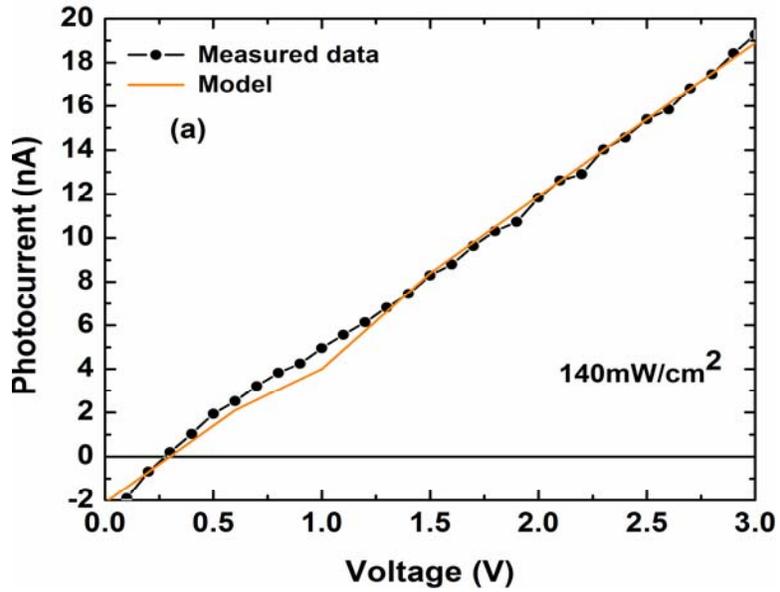

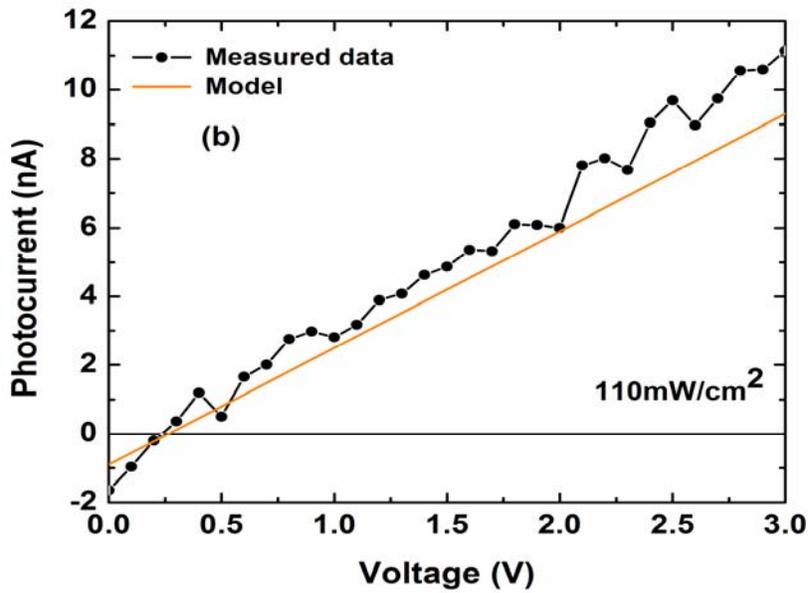

**Figs 5.7.** Comparison of I-V data: predicted model data compared with the experimental data for the PLZT PV device under irradiance of a) 140mW/cm$^2$ and (b) 110mW/cm$^2$.

The experimentally measured data for same irradiance is also represented in plots for comparison. It is observed from Fig. 7(a) that for the I-V data measured under 140 mW/cm$^2$, the model results fits the measured data with a coefficient of determination ($R^2$) [24] value of 0.9978, whereas for the data of 110mW/cm$^2$ from Fig.7 (b), model fits with $R^2$ of 0.9860. It is to be noted that the model fits better for the measured data at higher irradiance ($R^2$ of the model is closer to 1 at 140 mW/cm$^2$), which could be attributed to more generated carriers at higher irradiance levels, thus suppressing the noise. Data for irradiance level higher than 140 mW/cm$^2$ could not be measured due to the limitation of our experimental setup.

## 5.4  PLZT Solar Cell Circuit Model

A cell electric circuit model of the PLZT PV cell based on equation (3) was developed. Equation (3) can be represented by a voltage-controlled current-source (VCCS), a current source (I_1) and an internal resistance (R) varied with irradiance. The circuit model developed based on equation (3) is shown in Fig. 5.8.

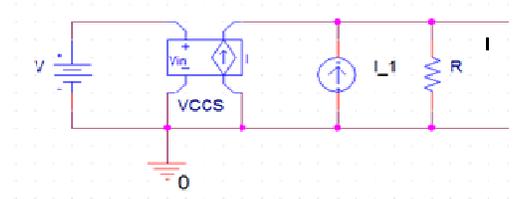

**Fig. 5.8.** PLZT electric circuit cell circuit model

In Fig. 5.8, V is the cell operating voltage. I_1 represents the term $(-4\times10^{-11}\cdot G + 3.5\times10^{-9})$ from equation (3), which is dependent on irradiance (G) but independent of V. A voltage controlled current source (VCCS) represents the term $(1.62\times10^{-10}\cdot G - 9.8\times10^{-9})\cdot V$ from equation (3). The value of internal resistance R varies with irradiance and is calculated as:

$$R = \frac{V}{I} = \frac{V}{f(V)} \qquad (4)$$

where $f(V)$ is determined by equation (3).

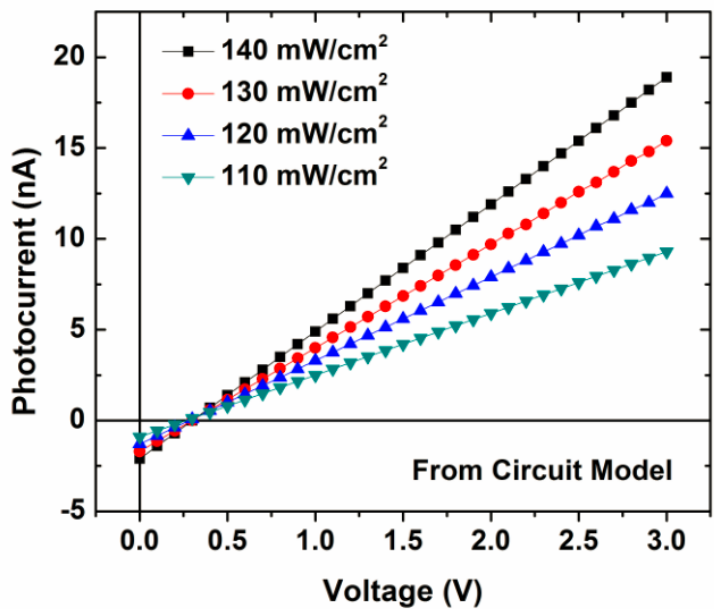

**Fig. 5. 9.** I-V curves of PLZT photovoltaic cell circuit model under various irradiance levels

Fig. 5.9 shows the I-V curves derived from the circuit model (Fig. 8) at various irradiance levels from 110-140 mW/cm$^2$. It can be observed that, the simulated data from Fig. 9 have a close match when compared with experimental data in Fig. 2. This data was obtained by simulating the electric circuit model in Pspice® electric circuit simulation software.

In this paper we have used an equivalent circuit model to describe a single ferroelectric photovoltaic cell. We believe that development of such circuit model is significant and it could be used further to predict the I-V behavior of multiple ferroelectric PV devices, when connected in series and/or in parallel, which is a part of future work.

## 5.5    Conclusion

A mathematical model suitable for a ferroelectric based photovoltaic device using varying irradiance was developed using a Simulink® tool of MATLAB®. The model was used to characterize a single PV cell I-V data, by choosing inputs such as irradiance and the cell voltage. The output of the model obtained from the simulation was used to predict the electrical behavior of ferroelectric photovoltaic cells for a range of irradiance and cell voltage values. Additionally, based on mathematical analysis and the experimental data, an equivalent circuit model of the cell was developed. The model in this work has been developed to predict the change in the photocurrent with changing irradiance and cell voltage for PLZT based ferroelectric device. The proposed model is limited to the fabricated PLZT based ferroelectric photovoltaic device with dimensions $9.1 \times 10^{-8}$ $m^2$. However, the same modeling procedure can be used to for any ferroelectric device with different dimensions. In that case, the co-efficient for the equation used in the model would change and need to be updated. This model also can be extended for multiple ferroelectric device structures, connected in series and/or in parallel. Future work will shed more light on this aspect.

**5.6  Acknowledgements**

Authors H.V.N. and S.K. would like to acknowledge National Science Foundation under ECCS Grant No. 0943711 and Research Grants Committee of The University of Alabama for support of this work. Y. J. and J.A.Q. would like to acknowledge the support of the Research Grants Committee of The University of Alabama. Any opinions, findings and conclusions or recommendations expressed in this material are those of the author(s) and do not necessarily reflect the views of the funding agency or source.

## 5.7 Chapter 5 References

1    S. Y. Yang, J. Seidel, S. J. Byrnes, P. Shafer, C. H. Yang, M. D. Rossell, P. Yu, Y. H. Chu, J. F. Scott, J. W. Ager, L. W. Martin, and Re. Ramesh, "Above-bandgap voltages from ferroelectric photovoltaic devices", Nature Nanotechnology, 5 (2), 2010 p. 143-147.

2    M. Qin, K. Yao, Y. C. Liang, and B. K. Gan, "Stability of photovoltage and trap of light-induced charges in ferroelectric $WO_3$ doped $(Pb_{0.97}La_{0.03})(Zr_{0.52}Ti_{0.48})O_3$ thin films", Applied Physics Letters, 91 (9) 2007 p. 64562.

3    H. T. Huang, "Solar energy: Ferroelectric photovoltaics, Nature Photonics", 4 (3), 2010, p. 134-135.

4    G. Bee Keen, Y. Kui, L. Szu Cheng, C. Yi Fan, and G. Phoi Chin, "An Ultraviolet (UV) Detector Using a Ferroelectric Thin Film With In-Plane Polarization", Electron Device Letters, IEEE, Vol. 29 (11), 2008, p.1215-1217.

5    G. Bee Keen, Y. Kui, L. Szu Cheng, G. Phoi Chin, and C. Yi Fan, "A Ferroelectric Ultraviolet Detector With Constructive Photovoltaic Outputs", Electron Device Letters, IEEE, 32 (5), June 2011, p.665-667.

6    M. Qin, K. Yao, and Y. C. Liang, "High efficient photovoltaics in nanoscaled ferroelectric thin films", Applied Physics Letters, 93 (12), 2008, p. 122904-122903.

7    M. Qin, K. Yao, and Y. C. Liang, "Photovoltaic mechanisms in ferroelectric thin films with the effects of the electrodes and interfaces", Applied Physics Letters, 95 (2), April 2009, p. 022912-022913.

8    R. Bharanikumar, A. Nirmal Kumar, K.T. Maheswari, "Novel MPPT controller for wind turbine driven permanent magnet generator with power converters", The Journal of International Review of Electrical Engineeing, Vol. 5(4), 2010, p.1555-1562.

9    M. Shafiee, E.S. Jafarabadi, "Dynamic performance of wind/PV/battery/fuel-cell hybrid energy system", International Review of Modelling and Simulations (I.RE.MO.S.), Vol. 2 (1), February 2009,p. 25-34.

10   T.T.N. Khatib, A.Mohamed, M. Mahmoud, N. Amin, "An efficient maximum power point tracking controller for a standalone photovoltaic system", International Review on Modelling and Simulations (I.RE.MO.S.), Vol. 3 (2), 2010,p. 129-139.

11   Y. Jiang and J. Abu Qahouq, "PV System Matlab Simulation Model for Two MPPT Methods", International Review on Modelling and Simulations (I.RE.MO.S.), 3 (5), 2010. p. 1002-1009.

12   R. Ramaprabha, N. Balamurugan and B.L. Mathur, "Implementation of particle swarm optimization based maximum power point tracking of solar photovoltaic array under non uniform insolation conditions", International Review on Modelling and Simulations (I.RE.MO.S.), 6 (3), 2011.p. 1503-1510.

13   R. Chenni, L. Zarour, M. Amarouayache and A. Bouzid, "A new design for analogue maximum power point tracking" ,International Review on Modelling and Simulations (I.RE.MO.S.), 3 (1), 2010, p. 93-99.

14   H. Abouobaida, M. Cherkaoui and M. Ouassaid, "Robust maximum power point tracking for fast changing environmental conditions", International Review on Modelling and Simulations (I.RE.MO.S.), 4 (1), , 2011.p. 391-396.

15   M. Taherbaneh, A.H. Rezaie, H. Ghafoorifard, M.B. Menhaj and J.M. Milimonfared, "Improving the PV panel model based on manufacturer data" , International Review on Modelling and Simulations (I.RE.MO.S.), 3 (4), 2010.p. 621-629.

16   A. Blorfan, D. Flieller, P. Wira, G. Sturtzer and J. Mercke, "A new approach for modeling the photovoltaic cell using orcad comparing with the model done in Matlab", International Review on Modelling and Simulations (I.RE.MO.S.), 3 (5), 2010.p. 948-954.

17   Y. Jiang, J. Abu Qahouq, M. Orabi, M.El-Sayed, and A. Hassan, "Energy Efficient Fine-grained Approach for Solar Photovoltaic Management System", The IEEE International Telecommunications Energy Conference (INTELEC), October 2011.

18   M. Qin, K. Yao, and Y. C. Liang, "Photovoltaic characteristics in polycrystalline and epitaxial $(Pb_{0.97}La_{0.03})(Zr_{0.52}Ti_{0.48})O_3$ ferroelectric thin films sandwiched between different top and bottom electrodes", Journal of Applied Physics, 105 (6) 2009. p. 648746.

19   C. Bin, L. Mi, L. Yiwei, Z. Zhenghu, Z. Fei, Z. Qing-Feng, and L. Run-Wei, "Effect of top electrodes on photovoltaic properties of polycrystalline $BiFeO_3$ based thin film capacitors", Nanotechnology, 22 (19), 2011.p. 195201.

20   R. P. Vengatesh and S. E. Rajan, "Investigation of cloudless solar radiation with PV module employing Matlab-Simulink", Emerging Trends in Electrical and Computer Technology (ICETECT), 2011, p. 141-147.

21   J. A. Gow and C. D. Manning, "Development of a model for photovoltaic arrays suitable for use in simulation studies of solar energy conversion systems", Sixth International Conference on Power Electronics and Variable Speed Drives, 429 , September 1996, p. 69-74.

22  J. A. Gow and C. D. Manning, "Development of a photovoltaic array model for use in power-electronics simulation studies",  Electric Power Applications,146, 1999, pp. 193-200.

23  V. N. Harshan and S. Kotru, "Effect of Annealing on Ferroelectric Properties of Lanthanum Modified Lead Zirconate Titanate Thin Films", Integrated Ferroelectrics, Vol. 130 (1), 2011, p. 73-83.

24  N. R. Draper and H. Smith, "Applied Regression Analysis" (Wiley-Interscience, 1998)

# CHAPTER 6

# EVALUATION OF ITO FILMS GROWN AT ROOM TEMPERATURE BY PULSED ELECTRON DEPOSITION [5]

**Abstract**

Good quality Sn-doped $In_2O_3$ films, with thickness of 30 nm were deposited using a vapor deposition technique known as pulsed electron deposition (PED). The films were deposited on (100) Si substrates, at room temperature from a ceramic target of ITO (90/10). A pulsed electron beam was used for ablation of the target. Voltage of the electron source was maintained at 18 KV with frequency of pulses at 3 Hz. Distance between source and target was maintained around 6 mm, and the substrate to target distance was ~ 7cm. Oxygen pressure in the chamber during growth was varied from 3.1 mTorr to 20 mTorr. To evaluate the quality of grown films, various characterization techniques were employed. Effect of oxygen chamber pressure on resistivity ($\rho$), carrier concentration (N), mobility ($\mu$) and optical constants (n & k) was carried out. Optical transparency and electrical conductivity of the films was seen to improve with increasing oxygen pressure. Details about the film preparation and evaluation of properties are reported.

---

[5] The work in this chapter has been published as "**Evaluation of indium tin oxide films grown at room temperature by pulsed electron deposition**" Harshan V. Nampoori, Veronica Rincon, Mengwei Chen, and Sushma Kotru,, J. Vac. Sci. Technol. A **28**, (2010), 671.

**6.1 Introduction**

Transparent conducting oxides (TCO) are the materials which combine high transparency in the visible region with high electrical conductivity. Such properties make TCO's suitable candidate material for a wide variety of applications, ranging from optoelectronic devices to architectural uses. Among the various TCOs, oxides of Indium (In), Tin (Sn) and Zinc (Zn) have been the focus of research in the past decade[1,2]. This work is focused on growth and evaluation of tin doped indium oxide, commonly referred to as ITO. ITO is an n-type semiconductor which has a band gap of 3.7 to 4.5 eV. ITO exhibits interesting electronic and optical properties including high optical transmittance (> 90 %) from visible to near-infra red wavelengths and low values of resistivity ($10^{-4}$ Ohm-cm). Thin films of ITO find use in flat panel displays, liquid crystal displays, organic light emitting diodes (OLEDS), photovoltaic devices (solar cells), thin film transistors, sensors, energy efficient windows and others [3].

ITO is formed by doping $In_2O_3$ with Sn. The conductivity is a direct consequence of formation of either SnO ($2^+$ valence of tin) or $SnO_2$ ($4^+$ valence of tin) during the doping process. The oxygen vacancies also play an important role in conduction mechanism. Major research efforts are focused towards obtaining ITO films to achieve highest possible mobility with lowest resistivity while maintaining the transparency of the films. There are presently many physical deposition techniques available to obtain high quality ITO films such as sputtering, e-beam evaporation and PLD [4-8]. Among the chemical deposition techniques, spray pyrolysis and dip coating are reported to result in good quality films of ITO [9]. ITO films grown by sputtering technique meet the current requirements and quality for device applications; however to achieve such good quality films, the films are either grown at elevated substrate temperatures or require post annealing. This work explores a relatively newer vapor deposition technique known as pulsed electron deposition (PED) for the growth of ITO films where the films are deposited at room temperature.

Conceptually PED is similar to the pulsed laser deposition method (PLD) which is a technique of choice for depositing oxide materials including ITO [7,8,10]. In PLD short pulses of photons (generated using a laser beam) are used to ablate the ceramic source target. The major advantage of this method is its ability to maintain the stoichiometry of the films with respect to target [8,10]. However, limitations of this technique include high cost of laser source, eye-safety requirements, as well as issues related to scalability. In PED technique instead of photons, energetic electrons are used to ablate the ceramic target, which is suitable for wide band gap materials. This process being scalable and low cost, makes this technique worth exploring for depositing films of various materials including TCOs.

In this article, growth of ITO films by PED and the preliminary results on the optical and electronic properties of these films is reported. Previous work on growth of TCO films by PED includes deposition of $SnO_2$ and Mo doped $In_2O_3$ films [11,12]. To this best of our knowledge, this is the first report on ITO films grown by PED. The motivation for this work is the integration of ITO films with ferroelectric materials for photovoltaic applications.

## 6.2 Experimental

ITO films were deposited on (100) Si substrates using pulsed electron deposition (PED) technique [11]. In this method an electron gun is used as a source for generating a pulsed electron beam. The PED gun used in our experiments is a commercial source, PEBS-20, manufactured by Neocera, Inc. The PED gun parameters are: 8-20 KV charging potential, 3-20 mTorr of background gas pressure, beam energy of 0.2-0.8 J, pulse width of 100 ns, maximum power density of $1.3 \times 10^8$ W/cm$^2$, a minimum beam cross section of $6 \times 10^{-2}$ cm$^2$ and a rep rate of up to 10 Hz. A schematic of the chamber used for depositing the films is shown in Fig. 6.1 [13]. Here the electron beam acts as a source to generate plasma from a ceramic target. The target used is a commercially obtained sintered ceramic ITO target (90 wt % $In_2O_3$+10 wt % $Sn_2$), one inch diameter and 0.25 inch thick.

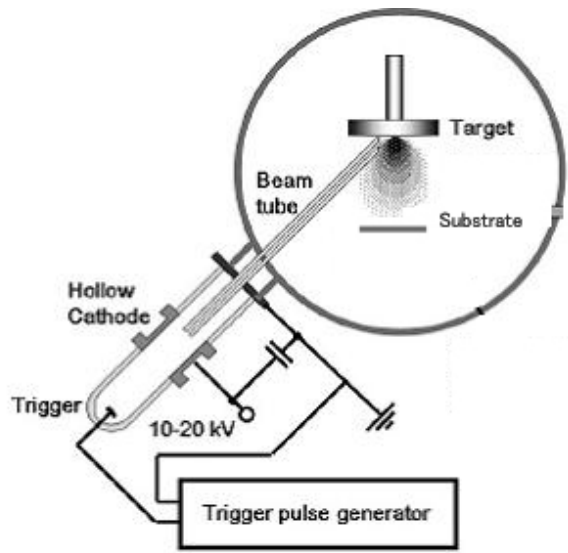

**Figure 6.1** Schematic diagram of a pulsed electron deposition system used for deposition of ITO films

During all our experiments, the voltage of the electron source was maintained at 18 KV with frequency of pulses at 3 Hz. Distance between source and target was maintained around 6 mm. The target to the substrate distance was fixed at 7 cm. The vacuum chamber was evacuated to the base pressure of 2 X $10^{-6}$ Torr prior to deposition. Films were deposited at four pressures viz., 3.1, 6.5, 15 and 20 mTorr, the pressure being maintained by controlling the flow of oxygen gas (99.999%, pure) introduced in the chamber. The substrate temperature was maintained at room temperature. All the samples were deposited with 5,000 pulses resulting in films thickness of 30 nm and a deposition rate of 0.06A/pulse.

The sheet resistance ($R_s$) of films was measured using a four-point probe (Jandel, RM3) Resistance was measured at 5 different points on the films, and then averaged over these values. Films thickness was measured by profilometry (Dektak IIA) and verified by spectroscopic ellipsometry (VASE, JA Wollam). Resistivity of these films was calculated from the relation, $\rho = R_s t_s$, where $t_s$ is the sample thickness. Here sample thickness was assumed to be uniform through the film. Resistivity values were averaged over three measurements at each point. Mobility and carrier concentration of these films were measured using four probe resistivity setup (MMR Technologies) under field strength of 2500 Oe provided by a permanent magnet. Data obtained from ellipsometry was used to calculate the optical constants such as refractive index (n) and absorption coefficient (k). The structural properties of these films were analyzed using X-ray diffraction (Rigaku Instruments).

## 6.3 Results and Discussion

### 6.3.1 Dependence of electrical properties on oxygen pressure

The electrical properties of ITO films are known to be depending on the deposition parameters such as substrate deposition temperature, oxygen pressure, film thickness and composition. During this work, the films were deposited at a constant substrate temperature (room temperature). Thickness of all the films was in the range of ~ 30 nm. Oxygen pressure was varied from 3.1 to 20 mTorr to observe the effect on the electrical properties of ITO films. Fig. 6.2 illustrates the variation of resistivity ($\rho$), carrier concentration ($N$), and Hall mobility ($\mu$) as a function of oxygen pressure for the ITO films grown at room temperature. It is found that the films grown at the lowest oxygen pressure of 3.1 mTorr exhibit the highest resistivity of $1.947 \times 10^{-3}$ ohm cm. As the pressure increases, the resistivity of the films drops. The lowest resistivity values ($0.226 \times 10^{-3}$ ohm cm) were seen for films grown at higher pressures (20 mTorr). Further increase in oxygen pressure was not possible due to the limitation of the PED process range.

**Table 6.1:** Resistivity, carrier concentration and mobility of ITO films

| Films | Thickness | Oxygen Pressure (mTorr) | Resistivity ($10^{-3}$ Ω-cm) | Carrier concentration ($cm^{-3}$) X $10^{20}$ | Mobility ($cm^2V^{-1}sec^{-1}$) |
|---|---|---|---|---|---|
| | | 3.1 | 1.947 | 0.0305 | 70.57244 |
| | | 6.5 | 0.4487 | 1.04 | 19.79761 |
| ITO/Si | 30 nm | 15 | 0.2282 | Not measured | Not measured |
| | | 20 | 0.2260 | 0.236 | 14.84361 |

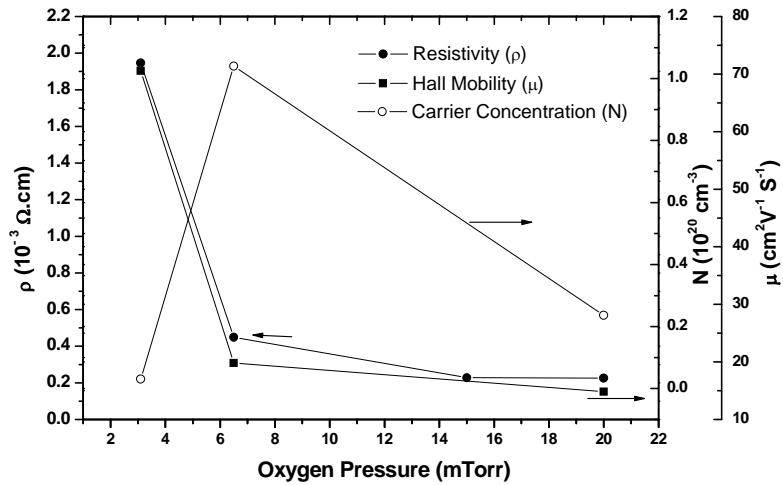

**Figure 6.2.** Variation of resistivity, carrier density, and Hall mobility as a function of oxygen deposition pressure for the films grown at room temperature. The film thickness was 30 nm for all films.

Films which were grown at the lowest pressure (3.1 mTorr) exhibit high value of mobility but a low carrier density, which results in high resistivity. It has been reported [14] that the factors influencing mobility are the grain boundaries and the scattering effect, due to the creation of oxygen vacancies. Crystallinity of the film and the relaxation time of the electron transport also affect the resistance of the sample [14]. As the pressure increases from 3.1 mTorr to 6.5 mTorr, mobility of the films drop whereas the carrier concentration goes up. When the oxygen pressure is increased further to 20 mTorr, there is no appreciable change in mobility whereas the carrier concentration of the films drops slightly. This indicates that higher pressure (3.1-20mTorr) leads to incorporation of oxygen into the films creating less $O_2$ vacancies thereby reducing mobility, increasing carrier concentration and thus reducing the resistivity.

The quantitative values of resistivity, carrier concentration and mobility of the films grown at various oxygen pressures are summarized in Table 6.1. A pressure range of 6-20 mTorr is suitable for obtaining films by PED, with low values of resistivity, and high values for mobility and carrier concentration. Although the resistivity of the films is still high compared to some reported values, the combination of low resistivity coupled with high mobility and carrier concentration at room temperature deposited ITO films, makes the PED technique worth exploring for such work.

### 6.3.2. Dependence of optical properties on oxygen pressure

Optical constants such as refractive index and absorption co-efficient were calculated from data obtained using spectroscopic ellipsometer. Fig. 6.3 shows the dependence of refractive index (n), and extinction coefficient (k) on deposition pressure for ITO films grown at room temperature. As observed from Fig. 6.3, films deposited at lower pressure have refractive index close to 2.0 (at 550 nm) where as the films deposited at higher pressure show lower value for n (<2.0) at the same wavelength. Thus as the working pressure increases, the refractive index and the extinction coefficient decreases. A similar behavior has been reported for the sputtered ITO thin films [7]. The optical constants of the commercially available ITO films grown on glass substrates are also reported to vary with annealing and roughness of the films [15]. Such studies are underway for the PED grown films and will be reported in future communications.

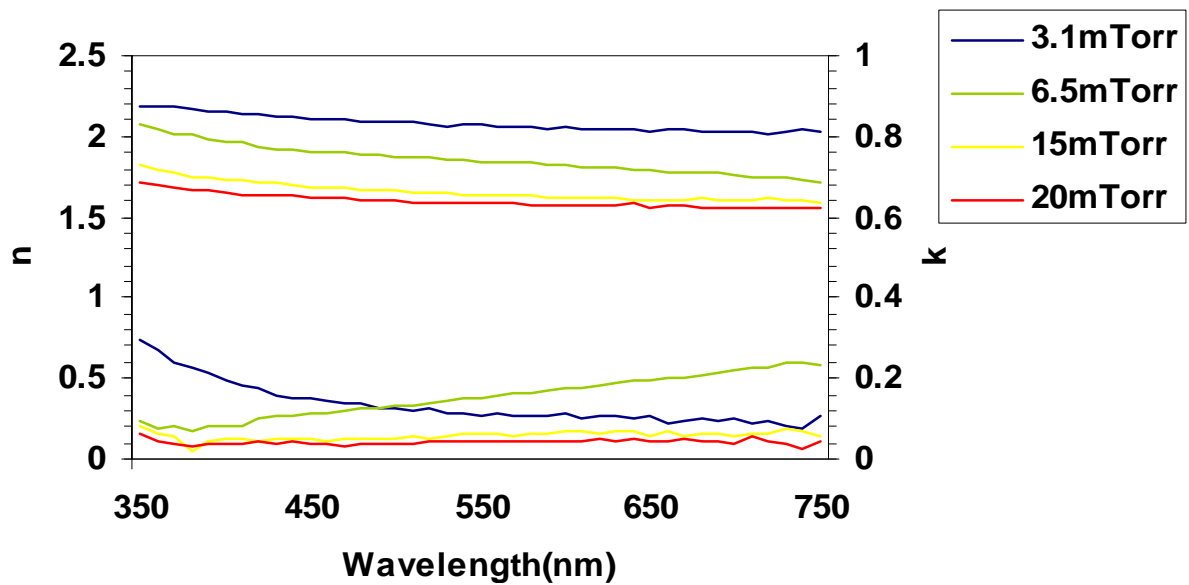

**Figure 6.3**.  Variation of refractive index (n) and extinction coefficient (k) for ITO films grown at different deposition pressures

### 6.3.3. Structural properties

Fig. 6.4 (a) shows the x-ray diffraction pattern of a 30 nm thin film of ITO deposited on Si (100); the film was deposited at room temperature at two oxygen pressures. It is clear that the phase of ITO films does not change when grown under different oxygen pressures as both films show strong peaks from (222), (411) and (431), which correspond to the ITO peaks (data compared with JCPDS files for ITO). Fig. 6.4 (b) shows x-ray diffraction patterns for an unannealed and annealed film, the film being annealed at 400°C for 40 minutes in flowing oxygen.

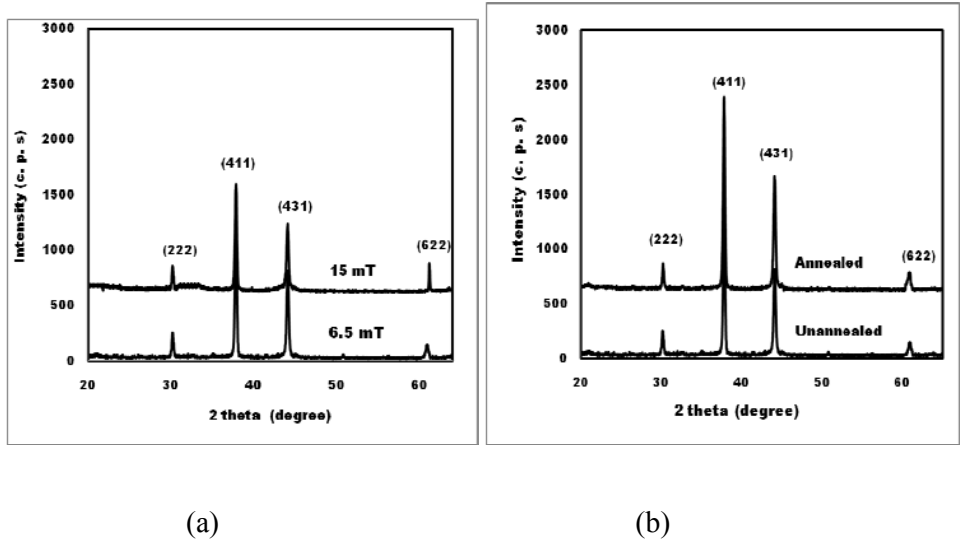

(a)          (b)

**Figure 6.4**  XRD diffraction patterns of ITO films deposited on Si (100), with thickness of 30 nm (a) films deposited at two pressures (b) unannealed and annealed films

The optimized annealing temperature and time were obtained by annealing the films at various temperatures and times (not reported here). The post annealed films do not show any change or shift in peaks or in the phase. This suggests that as deposited films by PED under the reported conditions are crystalline and further annealing of films does not help in promoting the crystallinity.

## 6.4 Conclusion

Thin films of ITO with thickness ~30 nm were grown on silicon wafers at room temperature using pulsed electron deposition technique. Effect of various oxygen pressures on the electrical and optical properties was investigated. Films with low resistivity values ~4.487X $10^{-4}$ ohm cm, mobility ~19.79761 $cm^2$ $V^{-1}sec^{-1}$ and carrier concentration ~1.04 X $10^{20}$ $cm^{-3}$ were achieved when films were deposited at 6.5 mTorr oxygen pressure. These results suggest that films grown by PED at room temperature in oxygen pressure range of 6-20 mTorr is the optimum range for obtaining films with relatively low resistance, high mobility and carrier concentration.

## 6.5 Acknowledgements

The work was supported by the National Science Foundation under ECCS Grant No. 0943711. The authors extend their thanks to the Center for Materials and Information Technology (MINT) and Central Analytical Facility (CAF), University of Alabama, for use of characterization facilities. The PED unit was procured from startup funds provided by college of engineering at UA.

## 6.6 Chapter 6 References

1     T. Minami, in J. Vac. Sci. Technol. A (1999), p. 1765-1772 .

2     Semiconducting Transparent Thin Films, D. a. L. Hartnagel H L, Jain a K and Jagadish C, Institute of Physics Publishing, 1995

3     C. G. Granqvist, "Transparent conductors as solar energy materials: A panoramic review," Solar Energy Materials and Solar Cells, 91 (17), (2007), p. 1529-1598.

4     C.-H. Yang, S.-C. Lee, T.-C. Lin, and S.-C. Chen, "Electrical and optical properties of indium tin oxide films prepared on plastic substrates by radio frequency magnetron sputtering," Thin Solid Films, 516 (8), (2008), p. 1984-1991.

5     M. Bender, W. Seelig, C. Daube, H. Frankenberger, B. Ocker, and J. Stollenwerk, "Dependence of oxygen flow on optical and electrical properties of DC-magnetron sputtered ITO films," Thin Solid Films, 326 (1-2), (1998), p. 72-77.

6     H. R. Fallah, M. Ghasemi, A. Hassanzadeh, and H. Steki, "The effect of annealing on structural, electrical and optical properties of nanostructured ITO films prepared by e-beam evaporation," Materials Research Bulletin, 42 (3), (2007), p. 487-496.

7     H. Kim, C. M. Gilmore, A. Pique, J. S. Horwitz, H. Mattoussi, H. Murata, Z. H. Kafafi, and D. B. Chrisey, "Electrical, optical, and structural properties of indium--tin--oxide thin films for organic light-emitting devices," Journal of Applied Physics, 86 (11), (1999), p. 6451-6461.

8     J. P. Zheng and H. S. Kwok, "Low resistivity indium tin oxide films by pulsed laser deposition," Applied Physics Letters, 63 (1), (1993), p. 1-3.

9     S. M. Rozati and T. Ganj, "Transparent conductive Sn-doped indium oxide thin films deposited by spray pyrolysis technique," Renewable Energy, 29 (10), (2004), p. 1671-1676.

10     C. Viespe, I. Nicolae, C. Sima, C. Grigoriu, and R. Medianu, "ITO thin films deposited by advanced pulsed laser deposition," Thin Solid Films, 515 (24), (2007), p. 8771-8775.

11     R. J. Choudhary, S. B. Ogale, S. R. Shinde, V. N. Kulkarni, T. Venkatesan, K. S. Harshavardhan, M. Strikovski, and B. Hannoyer, "Pulsed-electron-beam deposition of transparent conducting $SnO_2$ films and study of their properties," Applied Physics Letters, 84 (9), (2004), p. 1483-1485.

12     X. F. L. Li Huang, Qun Zhang, a Wei-Na Miao, Li Zhang, Xue-Jian Yan, and A. Z.-Y. H. Zhuang-Jian Zhang, in J. Vac. Sci. Technol. A (Sep/Oct 2005), p. 135-1353.

13  H. M. Christen, D. F. Lee, F. A. List, S. W. Cook, K. J. Leonard, L. Heatherly, P. M. Martin, M. Paranthaman, A. Goyal, and C. M. Rouleau, "Pulsed electron deposition of fluorine-based precursors for $YBa_2Cu_3O_{7-x}$-coated conductors," Superconductor Science & Technology, 18 (9), (2005), p. 1168-1175.

14  S. Calnan and A. N. Tiwari, "High mobility transparent conducting oxides for thin film solar cells," Thin Solid Films, In Press, Corrected Proof.

15  S. D'elia, N. Scaramuzza, F. Ciuchi, C. Versace, G. Strangi, and R. Bartolino, "Ellipsometry investigation of the effects of annealing temperature on the optical properties of indium tin oxide thin films studied by Drude-Lorentz model," Applied Surface Science, 255 (16), (2009), p. 7203-7211.

# CHAPTER 7

# CONCLUSION AND FUTURE WORK

This dissertation investigates photovoltaic properties of ferroelectric $Pb_{0.95}La_{0.05}Zr_{0.54}Ti_{0.46}O_3$ thin films. The films were prepared using solution based metal-organic decomposition route and spin coating method. Multiple experiments were required to obtain the parameters which resulted in optimized ferroelectric properties.

The photovoltaic characteristics of the grown films were investigated using a capacitor type device. An experimental measurement setup was assembled to measure the PV properties. Post annealing temperatures of these films were found to influence the photo response. The film annealed at $750^oC$ was found to show the highest photocurrent. The effect of illumination on the properties such as capacitance, ferroelectric polarization and leakage current was analyzed. The capacitance and the polarization of the PLZT films were observed to be suppressed on illumination.

The electric field at the metal-ferroelectric interface and the depolarization field within the films is observed to affect the photovoltaic characteristics. Two types of electrodes with different work functions were investigated to study the PV response of the capacitor type devices. It was shown that the use of a low work function metal electrode, such as aluminium, increases the efficiency of the ferroelectric photovoltaic devices.

The current–voltage (I-V) characteristics were measured and modeled as a function of irradiance. The current response at a given irradiance was modeled and found to fit to a line equation. An equivalent electric circuit model based on the characteristic equation was developed, to predict the behavior of a single PLZT photovoltaic cell under various levels of irradiance.

Pulsed electron deposition technique was investigated to prepare films of a transparent conducting material, ITO. The grown films were evaluated for resistivity ($\rho$), carrier concentration (N), mobility ($\mu$) and optical constants (n & k). Optical transparency and electrical conductivity of the PED grown films was seen to be comparable to the films prepared by other methods. The purpose of this work was to evaluate the feasibility of these films as top electrodes for the capacitor devices. Use of transparent conducting oxides is expected to further improve the efficiency of the devices.

The area of ferroelectric photovoltaic is relatively young, and more fundamental research needs to be carried out, before this technology can become a reality for future PV devices. In this work, some fundamental research issues such as optimization of process parameters to achieve films with good ferroelectric properties, is addressed which is essential to obtain a good PV response. The influence of low work function electrodes was investigated as a means to improve the efficiency of the devices. The focus of this work was on polycrystalline PLZT films. Further research is needed to study the photovoltaic properties of epitaxial films, which are expected to improve the efficiency further. The growth of PZT films on a $PbTiO_3$ template has shown to improve the ferroelectric properties. Improving the orientation of ferroelectric films by introducing a template layer is another way to improve the efficiency of these devices, and should be studied.

The ferroelectric-electrode interfaces often act as traps for oxygen vacancies and are not neutral. The charges generated in bulk are trapped at these interfaces and do not contribute to the total photocurrent from the material. The use of conducting oxide electrodes can significantly reduce the number of oxygen vacancy traps. Thus, proper choice of the electrodes can further improve the efficiency of the devices and should be explored.

The model established for single ferroelectric cell can be extended to multiple cells connected together to improve the power output of these devices. For such model, it is crucial to have a control on the internal field of each cell. This work should be pursued further to improve the device efficiency.

# APPENDIX

# List of Publications

**JOURNAL PAPERS**

1. **Photovoltaic and ferroelectric properties of $Pb_{0.95}La_{0.05}Zr_{0.54}Ti_{0.46}O_3$ thin films under dark and illuminated conditions,** Harshan V N and Sushma Kotru submitted to Journal of Ferroelectrics, May 2012

2. **Irradiance dependent equivalent model for PLZT based PV devices,** Harshan V Nampoori, Jiang Yuncong, Jaber Abu Qahouq, Sushma Kotru, International Review on Modelling and Simulations (I.RE.MO.S), 5(1), 517, 2012.

3. **Influence of work-function of top electrodes on the photo-voltaic characteristics of $Pb_{0.95}La_{0.05}Zr_{0.54}Ti_{0.46}O_3$ thin film capacitors,** Harshan V. N and Sushma Kotru, Applied Physics Letters 100 (17), 173901, 2012.

4. **Electrical and optical properties of ITO films grown at room temperature on glass substrates using pulsed electron deposition,** Sushma Kotru, Mengwei Chen, Harshan V. Nampoori, Rachel M. Frazier, Journal of Optoelectronics and Advanced Materials 14, (1-2), 106, 2012.

5. **Effect of annealing on ferroelectric properties of lanthanum modified lead zirconate titanate thin films**, Harshan V. N and Sushma Kotru, Integrated Ferroelectrics 130 (1), 73, 2011.

6. **Evaluation of ITO films grown at room temperature by Pulsed Electron Deposition**, Harshan V. N, Veronica Rincon, Mengwei Chen and Sushma Kotru, J Vac. Science. Technol. A Volume 28, Issue 4, pp. 671-674, 2010.

7. **Effect of kinetic growth parameters on leakage current and ferroelectric behavior of $BiFeO_3$ thin films**, Vilas Shelke, Harshan V.N , Sushma Kotru and Arunava Gupta, Journal of Applied Physics, 106 104114, 2009.

**CONFERENCE PAPERS**

1.  **Feasibility studies of using PED deposited Sn-doped $In_2O_3$ films for organic electronic devices**, Sushma Kotru, Mengwei Chen, Harshan V. Nampoori and Rachel M Frazier, MRS Proceedings, 1327, mrss11-1327-g03-06, doi:10.1557/opl.2011.1119, 2011

2.  **Fabrication and magneto-capacitance measurements of $PbNb_{0.02}Zr_{0.2}Ti_{0.8}O_3/La_{0.7}Sr_{0.3}MnO_3/SiO_2/Si$ structure grown by chemical solution deposition**, Sushma Kotru and Harshan V. Nampoori, MRS Proceedings, 1161, 1161-I04-06, doi:10.1557/proc-1161-i04-06, 2009

**PRESENTATIONS**

1.  **PLZT for photovoltaic applications**, V N Harshan, Denzel Evans-Bell and Sushma Kotru, MRS, Spring review, SFO, CA, April 2011 (**Poster**)

2.  **Barium ferrite films for microwave applications,** Aaron, Harshan V N, Sushma Kotru, UA Undergraduate conference, UA, April 2011 (**Poster**)

3.  **Ferroelectric thin films for photovoltaic applications**, V N Harshan, Denzel Evans-Bell and Sushma Kotru, MINT annual fall review, MINT, UA, Nov 2010 (**Poster**)

4.  **Evaluation of transparent conducting Sn-doped $In_2O_3$ films deposited on glass substrates using Pulsed Electron Deposition**, Mengwei Chen, Harshan V. Nampoori, Veronica C. Rincon, Rachel M. Frazier and Sushma Kotru, MINT annual fall review, MINT, UA, Nov 2010 (**Poster**)

5.  **The photo induced ferroelectric properties of $Pb_{0.95}La_{0.05}Zr_{0.54}Ti_{0.46}O_3$**, Harshan V N and Sushma Kotru AVS 57 Albuquerque, NM, October 2010 (**Oral**)

6.  **The structure and characterization of ITO thin films on glass substrates** Mengwei Chen, Harshan V N and Sushma Kotru, AVS 57 Albuquerque, NM, October 2010 (**Poster**)

7.  **PV effects on ferroelectric $Pb_{0.95}La_{0.05}Zr_{0.54}Ti_{0.46}O_3$**, Harshan V N, Denzel E Bells and S Kotru, MINT Fall review 2010, October 2010 (**Poster**)

8.  **PLZT ferroelectric films for photovoltaic applications**, Harshan V and S. Kotru, MINT, UA, July 30, 2010 (**Oral**)

9. **The effect of substrate vicinity on ferroelectric properties of epitaxial BiFeO$_3$ thin films**, Vilas Shelke, V. N. Harshan, Sushma Kotru and Arunava Gupta, MRS Fall meeting, Dec 2009 **(Poster)**

10. **Evaluation of ITO Films Deposited at room temperature by Pulsed Electron Deposition**, Harshan V. Nampoori, Veronica Rincon, Mengwei Chen and Sushma Kotru, AVS-Meeting 56, San Jose, Nov 2009 (**Poster**)

11. **Fabrication of a Vibratory Gyroscope Based on Piezoelectric Actuators and Sensors using MEMS Technology,** Veronica Rincon, Harshan V N and Sushma Kotru, AVS-Meeting, San Jose, Nov 2009 (**Poster**)

12. **Construction of ultra high vacuum system for educational research**, Harshan V Nampoori and Gary Mankey, AVS-Meeting 56, San Jose Nov 2009 **( Poster presented for the competition-Student build vacuum systems)**

13. **Characteristics of SiO$_2$ thin films grown by Pulsed Electron Deposition technique**, Veronica Rincon, Harshan V. N, Mengwei Chen and Sushma Kotru, 13th Annual GSA Research and Thesis conference, UA, March 27, 2010 (**Poster**)

14. **Pb$_{0.95}$La$_{0.05}$Zr$_{0.42}$Ti$_{0.58}$O$_3$ Thin Films for Photovoltaic Applications**, Harshan V N, Sushma Kotru 13th Annual GSA Research and Thesis conference, UA, March 27, 2010 (**Oral**)

15. **Fabrication and Magneto-Capacitance of PZT/LSMO structure on SiO$_2$/Si**, Harshan V and S. Kotru, MINT, UA, Oct 30, 2009 (**Oral**)

16. **Chemical Solution deposition of LSMO thin films**, Harshan V. Nampoori and Sushma Kotru, International workshop on Multiferroics, UC Santa Barbara, CA, July 20-Aug 3$^{rd}$, 2008 (**Poster**)

17. **Fabrication and Magneto-Capacitance Measurements of PbNb$_{0.02}$Zr$_{0.2}$Ti$_{0.8}$O$_3$/La$_{0.7}$Sr$_{0.3}$MnO$_3$/SiO$_2$/Si Structure Grown by Chemical Solution Deposition,** Sushma Kotru and Harshan V. Nampoori, MRS April Symposium, SFO, April 2008 ( **Oral** )

18. **The magneto-electric memories for next generation**, Harshan V and S. Kotru, Graduate Student Research Council Presentation (GSRC), UA, March 2008 (**Oral**)